\documentclass[fleqn,usenatbib]{mnras}

\usepackage[T1]{fontenc}
\usepackage{ae,aecompl}
\usepackage{pdflscape}
\usepackage{comment}
\usepackage{subfigure}
\usepackage{multirow}
\usepackage{ulem}
\usepackage{caption}
\usepackage{multicol}

\usepackage{graphicx}	
\usepackage{amsmath}	
\usepackage{amssymb}	
\usepackage{bm}
\usepackage{pdflscape}

\usepackage{newtxtext,newtxmath}

\def\msun{\,{\rm M}_\odot}
\def\AU{\,{\rm AU}}
\def\zsun{\,{\rm Z}_\odot}
\def\dd{\,{\rm d}}
\def\d{{\rm d}}

\title[Evolution of massive stellar triples]{
Evolution of massive stellar triples and implications for compact object binary formation}

\author[J. Stegmann et al.]{Jakob Stegmann,$^{1}$\thanks{E-mail: StegmannJ@cardiff.ac.uk} Fabio Antonini,$^{1}$\thanks{E-mail: AntoniniF@cardiff.ac.uk} and Maxwell Moe$^{2}$\\
$^{1}$Gravity Exploration Institute, School of Physics and Astronomy, Cardiff University, Cardiff, CF24 3AA, UK\\
$^{2}$Steward Observatory, University of Arizona, 933 North Cherry Avenue, Tucson, AZ 85721, USA}

\date{Accepted 2022 July 29. Received 2022 July 6; in original form 2021 December 20}

\pubyear{2021}

\begin{document}
\label{firstpage}
\pagerange{\pageref{firstpage}--\pageref{lastpage}}
\maketitle

\begin{abstract}
Most back hole and neutron star progenitors are found in triples or higher multiplicity systems.
Here, we present a new triple stellar evolution code, ${\tt TSE}$, which simultaneously takes into account the physics of the stars and their gravitational interaction.
${\tt TSE}$ is used to simulate the evolution of massive stellar triples in the galactic field from the zero-age-main-sequence until they form compact objects.
To this end, we implement initial conditions that incorporate the observed high correlation between the orbital parameters of early-type stars.
We show that the interaction with a tertiary companion can significantly impact the evolution of the inner binary.
High eccentricities can be induced by the third-body dynamical effects, leading to a Roche lobe overflow or even to a stellar merger from initial binary separations $~10^3$~--~$10^5\,\rm R_\odot$.
In $\sim5\,\%$ of the systems the tertiary companion itself fills its Roche lobe, while $\sim 10\,\%$ of all systems become dynamically unstable.
We find that between $0.3\%$ and $5\%$ of systems form a stable triple with an inner compact object binary, where the exact fraction depends on metallicity and the natal kick prescription.
Most of these triples are binary black holes with black hole companions.
We find no binary neutron star in any surviving triple, unless zero natal kicks are assumed.
About half of all black hole binaries formed in our models are in triples, where in the majority the tertiary black hole can perturb their long-term evolution.
Our results show that triple interactions are key to a full understanding of massive star evolution and compact object binary formation.
\end{abstract}




\section{Introduction}
The majority of massive stars in the Galactic field is found to be confined in hierarchical triples in which a close inner binary is orbited by a distant outer companion \citep[][]{2017ApJS..230...15M,2015AJ....149...26A,2014ApJS..215...15S,2012Sci...337..444S}. Observations suggest that the triple star fraction amounts to roughly $\sim50\%$ and $70\%$ for early B-type and O-type stars, respectively \citep[][]{2017ApJS..230...15M}.
In general, hierarchical triples have been examined across a wide range of astrophysical scales, e.g., from satellites in low Earth orbits \citep[][]{2014AmJPh..82..769T}, planetary systems \citep[][]{2011Natur.473..187N,2013ApJ...779..166T,2015ApJ...799...27P,2015MNRAS.447..747L,2016ApJ...829..132P,2016MNRAS.456.3671A,2018AJ....155..118W,2018AJ....156..128S,2019MNRAS.484.5645V,2021ApJ...922....4S}, stellar black hole triples in the Galactic field \citep[][]{2017ApJ...836...39S,2018MNRAS.480L..58A,2017ApJ...841...77A,2018ApJ...863....7R,2021ApJ...907L..19V}, black hole triples in the dense cores of globular clusters \citep[][]{2002ApJ...576..894M,2003ApJ...598..419W,2014ApJ...781...45A,2016MNRAS.460.3494S,2019MNRAS.488.4370F,2021RNAAS...5...19R}, binaries near massive black holes \citep[][]{2012ApJ...757...27A,2016MNRAS.460.3494S,2018ApJ...856..140H,2019ApJ...878...58S,2021ApJ...917...76W}, to massive black hole triples at the centres of galaxies \citep[][]{2002ApJ...578..775B,2014MNRAS.439.1079A,2019MNRAS.486.4044B}. For sufficiently high inclinations, the secular gravitational perturbation from the tertiary companion leads to large-amplitude eccentricity oscillations in the inner binary, which are often referred to as Lidov Kozai (LK) oscillations \citep{1962P&SS....9..719L,1962AJ.....67..591K}.

Applied to massive stars, one can expect the presence of a  tertiary companion to enrich the variety of evolutionary pathways in the inner binary by driving it to close stellar interactions \citep[][]{2016ComAC...3....6T,2020A&A...640A..16T}. Yet, simulating the evolution of massive stellar triples poses a difficult challenge since the stellar physics of each individual star and the gravitational three-body dynamics have to be combined in a self-consistent way. Concerning massive stars, both of these aspects are closely intertwined. For instance, kicks experienced in the supernova (SN) explosions modify and potentially disrupt the three-body configuration. Likewise, massive stars at high metallicity suffer significant mass-loss through stellar winds that loosen the inner and outer orbits \citep[][]{1975ApJ...195..157C,1991A&A...252..159V,2001A&A...369..574V,2015ApJ...805...20S,2016MNRAS.459.3432M}. It has been shown that mass-loss in the inner binary due to winds or at compact object formation could induce or strengthen the LK effect \citep[][]{2013ApJ...766...64S,2014ApJ...794..122M}. In addition, massive stars attain large radii as they evolve off the main sequence and beyond, so that Roche lobe overflow and mergers are expected to occur frequently in isolated massive binaries \citep[][]{Bonnell:2005zp,2008MNRAS.384.1109E,2012Sci...337..444S,2021A&A...645A...5S}. For example, more than $\sim70\,\%$ of Galactic massive O-type stars are expected to undergo at least one mass-transfer episode with their binary companion \citep[][]{2012Sci...337..444S}. 
A tertiary companion could facilitate these types of close stellar interaction via the LK mechanism by driving the inner binary to smaller pericentre distances. Previous studies of stellar triples have shown that these may give rise to X-ray binaries \citep[][]{2016ApJ...822L..24N} or even trigger a stellar merger \citep[][]{2012ApJ...760...99P,2022arXiv220316544S} leading to the formation of Blue stragglers \citep[][]{2009ApJ...697.1048P,2014ApJ...793..137N,2016ApJ...816...65A} and type Ia SN \citep[][]{1999ApJ...511..324I,2011ApJ...741...82T}. Moreover, an expanded tertiary star could itself overflow its Roche lobe and initiate a mass transfer phase onto the inner binary \citep[][]{2014MNRAS.438.1909D,2019ApJ...876L..33P,2020MNRAS.491..495D,2020MNRAS.493.1855D,2020arXiv201104513H}.   

Modelling massive stellar triples will also help to understand the astrophysical origin of the binary mergers detected by LIGO and Virgo \citep[][]{2019PhRvX...9c1040A,2020arXiv201014527A,2021ApJ...913L...7A,2021arXiv211103634T}. It is unknown whether they resulted from isolated stellar evolution in which the binary stars harden during a common-envelope phase \citep[][]{2012ApJ...759...52D,2016Natur.534..512B,2018ApJ...856..140H,2018MNRAS.480.2011G,2021A&A...651A.100O} or via three-body interaction with a bound hierarchical companion \citep[][]{2017ApJ...836...39S,2017ApJ...841...77A,2018MNRAS.480L..58A,2018ApJ...863...68L,2018ApJ...863....7R,2020MNRAS.493.3920F,2021arXiv210501671M}, or they were driven by some dynamical interaction within a dense stellar environment, e.g., the dense cores of globular clusters \citep{2016PhRvD..93h4029R,2017MNRAS.469.4665P,2018ApJ...866L...5R,2020PhRvD.102l3016A}, massive young clusters \citep[][]{2010MNRAS.402..371B,2014MNRAS.441.3703Z,2019MNRAS.487.2947D,2021ApJ...913L..29F}, and galactic nuclei \citep[][]{2012ApJ...757...27A,2015ApJ...799..118P,2016ApJ...831..187A,2017ApJ...846..146P,2019ApJ...881L..13H,2020ApJ...894...15B,2021ApJ...917...76W}, or if the merger population formed in a combination of these channels \citep[][]{2021ApJ...910..152Z}. 

Several studies of the isolated triple channel incorporated stellar evolution, but the resulting binary black hole (BBH) mergers have yet to be studied self-consistently with stellar evolution. Particularly, the initial conditions of black hole triples are subject to large uncertainties since they elude direct detection. Studying the evolution of stellar triples from the zero-age-main-sequence (ZAMS) allows to derive parameter distributions of the black hole triples which are motivated by stellar observations. Thus, following the evolution of massive stellar triples is a preparatory work which is necessary for evaluating the isolated triple channel.

In this paper, we introduce the code {\tt TSE}\footnote{Publicly available at: \url{https://github.com/stegmaja/TSE}.} that  follows the secular evolution of hierarchical stellar triples from the ZAMS until they possibly form compact objects. {\tt TSE} builds upon the most updated prescriptions of the widely-adopted single and binary evolution codes {\tt SSE} \citep[][]{2000MNRAS.315..543H} and {\tt BSE} \citep{2002MNRAS.329..897H}, respectively, and employs the secular three-body equations of motion up to the octupole order with relativistic corrections up to the 2.5 post-Newtonian order. Thus, {\tt TSE} complements previous population synthesis codes which are designed to evolve stellar triples or higher-order configurations, e.g., {\tt MSE} \citep[][]{2020arXiv201104513H} and {\tt TrES} \citep[][]{2016ComAC...3....6T}. {\tt TSE} provides an evolution scheme for the stellar masses, radii, orbital elements, and spin vectors. We apply this code to a population of massive stellar triples to study their evolution until they form a double compact object (DCO) in the inner binary.

This paper is organised as follows. In Section~\ref{sec:methods}, we present our numerical framework. In the following sections, we apply it to a realistic population of massive stars. Section~\ref{sec:initial-conditions} motivates its initial conditions by current observations. In Section~\ref{sec:results}, we investigate different evolutionary pathways, present the final orbital distribution of triples that form compact objects, and discuss the impact of triple interactions on the evolution of massive stars. Finally, our findings are summarised in Section~\ref{sec:conclusions}.

If not stated differently, the magnitude, unit vector, and time derivative of some vector $\bm{V}$ are written as $V=\left|\bm{V}\right|$, $\bm{\hat{V}}=\bm{V}/V$, and $\bm{\dot{V}}=\d\bm{V}/\d t$, respectively. $G$ and $c$ refer to the gravitational constant and the speed of light, respectively. Coloured versions of the figures are available in the online journal.

\section{Method: Triple Stellar Evolution}\label{sec:methods}

\subsection{Triple dynamics}\label{sec:EoM}
In this section, we describe the numerical method we use to study the long-term evolution of hierarchical stellar triples. We are considering two stars with masses $m_{1(2)}$ and radii $R_{1(2)}$ that constitute an inner binary whose centre of mass is orbited by a distant outer stellar companion with mass $m_3$. The orbits are hierarchical in the sense that the inner semimajor axis is much smaller than the outer, i.e. $a_{\rm in}\ll a_{\rm out}$. The inner (outer) orbit carries an orbital angular momentum $\bm{L}_{\rm in(out)}$ with magnitudes
\begin{align}
    L_{\rm in}&=\mu_{12}\left[Gm_{12}a_{\rm in}\left(1-e_{\rm in}^2\right)\right]^{1/2},\label{eq:L_in}\\
    L_{\rm out}&=\mu_{123}\left[Gm_{123}a_{\rm out}\left(1-e_{\rm out}^2\right)\right]^{1/2},
\end{align}
where $m_{12}=m_1+m_2$ and $m_{123}=m_{12}+m_3$ are the total masses, $\mu_{12}=m_1m_2/m_{12}$ and $\mu_{123}=m_{12}m_3/m_{123}$ the reduced masses, and $e_{\rm in(out)}$ the eccentricity of the inner (outer) orbit. Furthermore, we define the mass ratios $q_{\rm in}=\min{(m_1,m_2)}/\max{(m_1,m_2)}$ and $q_{\rm out}=m_3/m_{12}$. 
The spatial orientation of the inner (outer) orbital frame can be characterised in terms of the dimensionless orbital angular momentum vector $\bm{j}_{\rm in(out)}$ and Laplace-Runge-Lenz vector $\bm{e}_{\rm in(out)}$ defined as 
\citep[e.g.,][]{2009AJ....137.3706T}
\begin{align}
    \bm{j}_{\rm in(out)}&=\sqrt{1-e_{\rm in(out)}^2}\bm{\hat{j}}_{\rm in(out)},\\
    \bm{e}_{\rm in(out)}&=e_{\rm in(out)}\bm{\hat{e}}_{\rm in(out)},
\end{align}
where $\bm{\hat{j}}_{\rm in(out)}$ and $\bm{\hat{e}}_{\rm in(out)}$ are unit vectors pointing along the orbital angular momentum $\bm{L}_{\rm in(out)}$ and the pericentre, respectively. Furthermore, the stars of the inner orbit spin carry some rotational angular momentum (spin) vector $\bm{S}_{1(2)}$ with magnitudes 
\begin{equation}\label{eq:S}
    S_{1(2)}=\kappa m_{1(2)}R_{1(2)}^2\Omega_{1(2)},
\end{equation}
where $\Omega_{1(2)}$ is the angular velocity of the rotating star and we set $\kappa=0.1$.

In this formalism, the secular equations of motion for the inner orbit, $\bm{j}_{\rm in}$ and $\bm{e}_{\rm in}$, its semimajor axis $a_{\rm in}$, and the spin vectors $\bm{S}_{1(2)}$ can be written as \citep[e.g.,][]{2016MNRAS.456.3671A} 
\begin{align}
    \frac{\dd\bm{j}_{\rm in}}{\dd t}&=\left.\frac{\dd\bm{j}_{\rm in}}{\dd t}\right\vert_\text{LK,Quad}+\left.\frac{\dd\bm{j}_{\rm in}}{\dd t}\right\vert_\text{LK,Oct}+\left.\frac{\dd\bm{j}_{\rm in}}{\dd t}\right\vert_\text{Tide}\nonumber\\
    &+\left.\frac{\dd\bm{j}_{\rm in}}{\dd t}\right\vert_\text{Rot}+\left.\frac{\dd\bm{j}_{\rm in}}{\dd t}\right\vert_\text{1.5PN}+\left.\frac{\dd\bm{j}_{\rm in}}{\dd t}\right\vert_\text{GW},\label{eq:j_A}\\
    \frac{\dd\bm{e}_{\rm in}}{\dd t}&=\left.\frac{\dd\bm{e}_{\rm in}}{\dd t}\right\vert_\text{LK,Quad}+\left.\frac{\dd\bm{e}_{\rm in}}{\dd t}\right\vert_\text{LK,Oct}+\left.\frac{\dd\bm{e}_{\rm in}}{\dd t}\right\vert_\text{Tide}\nonumber\\
    &+\left.\frac{\dd\bm{e}_{\rm in}}{\dd t}\right\vert_\text{Rot}+\left.\frac{\dd\bm{e}_{\rm in}}{\dd t}\right\vert_\text{1PN}+\left.\frac{\dd\bm{e}_{\rm in}}{\dd t}\right\vert_\text{1.5PN}+\left.\frac{\dd\bm{e}_{\rm in}}{\dd t}\right\vert_\text{GW},\label{eq:e_A}\\
    \frac{\dd a_{\rm in}}{\dd t}&=\left.\frac{\dd a_{\rm in}}{\dd t}\right\vert_\text{Tide}+\left.\frac{\dd a_{\rm in}}{\dd t}\right\vert_\text{Mass}+\left.\frac{\dd a_{\rm in}}{\dd t}\right\vert_\text{GW},\label{eq:a_in}\\
    \frac{\dd\bm{S}_{1(2)}}{\dd t}&=\left.\frac{\dd\bm{S}_{1(2)}}{\dd t}\right\vert_\text{Tide}+\left.\frac{\dd\bm{S}_{1(2)}}{\dd t}\right\vert_\text{Rot}+\left.\frac{\dd\bm{S}_{1(2)}}{\dd t}\right\vert_\text{Mass}\nonumber\\
    &+\left.\frac{\dd\bm{S}_{1(2)}}{\dd t}\right\vert_\text{1PN}\label{eq:S_1(2)},
\end{align}
where the terms on the r.h.s. are described in the following subsections. Treating the spins of the inner binary stars as vector quantities and including a vectorial prescription of tides, de Sitter precession, and Lense-Thirring precession (see below) is one main difference of {\tt TSE} compared to previous population synthesis codes.

\subsubsection{Eccentric Lidov-Kozai effect {\normalfont (LK)}}\label{sec:Lidov}
In a hierarchical configuration, the inner and outer orbit exchange angular momentum on secular timescales while separately conserving their orbital energies. As a consequence, the eccentricities $\bm{e}_{\rm in(out)}$ and directions of the orbital axes $\bm{\hat{j}}_{\rm in(out)}$ can oscillate in time while keeping the two semimajor axes $a_{\rm in(out)}$ constant \citep[LK,Quad;][]{1962P&SS....9..719L,1962AJ.....67..591K}.
Associated with these modes are well-defined minima and maxima for the mutual inclination $i=\cos^{-1}\bm{\hat{j}}_{\rm in}\cdot\bm{\hat{j}}_{\rm out}$ and the inner eccentricity $e_{\rm in}$. The oscillations between these extrema are the largest if the initial mutual inclination lies within the range of the so-called Kozai angles, $\cos^2 i<3/5$, and the timescale of these oscillations is given by \citep[e.g.,][]{2015MNRAS.452.3610A}
\begin{equation}\label{eq:t-LK}
    t_{\rm LK}\simeq\frac{1}{\omega_{\rm in}}\frac{m_{12}}{m_3}\left(\frac{a_{\rm out}j_{\rm out}}{a_{\rm in}}\right)^3,
\end{equation}
where $\omega_{\rm in}=\sqrt{Gm_{12}/a_{\rm in}^3}$ is the inner orbit's mean motion. Below, we will see that short-range forces between the tidally and rotationally distorted stars as well as relativistic effects cause the eccentricity vector $\bm{e}_{\rm in}$ to precess about the orbital axis $\bm{\hat{j}}_{\rm in}$. If this precession is sufficiently fast, the inner binary effectively decouples from the outer companion therefore suppressing any Lidov-Kozai oscillations \citep{2011Natur.473..187N,2015MNRAS.447..747L}. The critical value for $j_{\rm in}$ below which these are fully quenched can be found by requiring that the periapsis precesses faster about $\pi$ than $j_{\rm in}$ could change by the order of itself \citep{2018ApJ...863....7R}, i.e. by setting 
\begin{equation}\label{eq:timescales}
    \pi\min(t_{\rm Tide},t_{\rm Rot},t_{\rm 1PN})\leq j_{\rm in}t_{\rm LK},
\end{equation} 
with the associated timescales $t_{\rm 1PN}$, $t_{\rm Tide}$, and $t_{\rm Rot}$ defined in Eqs.~\eqref{eq:t-1PN}, \eqref{eq:t-Tide}, and \eqref{eq:t-Rot}, respectively.

In general, the quadrupole term provides a good approximation of the three-body dynamics only if the outer orbit is circular ($e_{\rm out}=0$). In order to properly describe systems with $e_{\rm out}>0$, we have to include the next-order octupole term (LK,Oct) in Eqs. \eqref{eq:j_A} and \eqref{eq:e_A} which is valid for non-axisymmetric outer potentials \citep[][]{2000ApJ...535..385F,2013MNRAS.431.2155N}. Compared to the quadrupole, this term is suppressed by a factor
\begin{equation}
    \epsilon_{\rm LK,Oct}=\frac{m_1-m_2}{m_{12}}\frac{a_{\rm in}}{a_{\rm out}}\frac{e_{\rm out}}{1-e_{\rm out}^2}.
\end{equation}
Following previous work, we neglect hexadecapole and higher-order terms in our models \citep[][]{2002ApJ...578..775B,2011Natur.473..187N,2013MNRAS.431.2155N,2018MNRAS.480L..58A,2018ApJ...863...68L,2018ApJ...863....7R,2019MNRAS.488.2480R,2021MNRAS.505.3681S,2021arXiv210501671M}. \citet[][]{2017PhRvD..96b3017W} showed that including the hexadecapole terms does not significantly alter the three-body dynamics unless the inner binary stars have near-equal masses in which extreme eccentricities could be achieved.  We used the timescales presented by \citet[][]{2021PhRvD.103f3003W} to estimate that in only $\sim3\,\%$ of the triples in our population (see Section~\ref{sec:initial-conditions}) the timescale of hexadecapole effects is shorter than the octupole and the typical stellar evolution timescale of massive stars ($\sim\rm Myr$). Hence, we opt for neglecting the hexadecapole terms. This introduces an uncertainty  that is arguably smaller than that due to the stellar evolution prescriptions.

We use Eqs.~(17)~--~(20) of \citet[][]{2015MNRAS.447..747L} for the LK contribution to the equations of motion of $\bm{e}_{\rm in}$ and $\bm{j}_{\rm in}$ which are used in {\tt TSE}.

\subsubsection{Tidal interaction {\normalfont (Tide)}}\label{sec:tides}
In close binaries, the mutual gravitational interaction between the stars raises tidal bulges on their surfaces \citep[e.g.,][]{1981A&A....99..126H,1989A&A...220..112Z,1998ApJ...499..853E}. The viscosity of the internal motion within the stars prevents these bulges to instantaneously align with the interstellar axis while dissipating kinetic energy into heat. Thus, the tilted tidally deformed stars torque each other leading to an exchange of rotational and orbital angular momentum. Generally, the strength of this interaction can be quantified in terms of a small lag time constant $\tau$ by which the tidal bulges lag behind or lead ahead the interstellar axis \citep[][]{1981A&A....99..126H}. In this work, $\tau$ is set to $1\,\rm s$ \citep[][]{2016MNRAS.456.3671A}. The full equations of motion for $\bm{e}_{\rm in}$, $\bm{S}_{\rm 1(2)}$, $a_{\rm in}$, and $\bm{j}_{\rm in}$ in Eqs.~\eqref{eq:j_A}~--~\eqref{eq:S_1(2)} (Tide) are adopted from Eqs.~(21), (22), and (56) of \citet[][]{2011CeMDA.111..105C}. Accordingly, the direction of the angular momentum flow and consequently the change of $e_{\rm in}$ and $a_{\rm in}$ depend on the ratio between orbital mean motion $\omega_{\rm in}$ and the spin rotation rate along the orbital normal $\Omega_{1(2)}\cdot\bm{\hat{j}}_{\rm in}$ \citep[][]{2011CeMDA.111..105C}
\begin{align}
    \dot{L}_{\rm in}&\propto\sum_{i=1,2}\left[\frac{f_5(e_{\rm in})}{j_{\rm in}^9}\frac{\bm{\Omega}_i\cdot\bm{\hat{j}}_{\rm in}}{\omega_{\rm in}}-\frac{f_2(e_{\rm in})}{j_{\rm in}^{12}}\right],\label{eq:L-Tide}\\
    \frac{\dot{e}_{\rm in}}{{e}_{\rm in}}&\propto\sum_{i=1,2}\left[\frac{11}{18}\frac{f_4(e_{\rm in})}{j_{\rm in}^{10}}\frac{\bm{\Omega}_i\cdot\bm{\hat{j}}_{\rm in}}{\omega_{\rm in}}-\frac{f_3(e_{\rm in})}{j_{\rm in}^{13}}\right],\label{eq:e-Tide}\\
    \frac{\dot{a}_{\rm in}}{{a}_{\rm in}}&\propto\sum_{i=1,2}\left[\frac{f_2(e_{\rm in})}{j_{\rm in}^{12}}\frac{\bm{\Omega}_i\cdot\bm{\hat{j}}_{\rm in}}{\omega_{\rm in}}-\frac{f_1(e_{\rm in})}{j_{\rm in}^{15}}\right],\label{eq:a-Tide}
\end{align}
where the polynomials $f_{1,2,...,5}(e_{\rm in})$ are given in Appendix~\ref{sec:Hut}. In our simulation, the initial rotational periods of the stars are typically a few days long, $1/\Omega_{1(2)}\sim\mathcal{O}(\rm days)$, which is much shorter or, at most, roughly equal to the initial orbital period (see Section~\ref{sec:initial-conditions}). Unless the stellar spins are retrograde ($\bm{\hat{\Omega}}_{1(2)}\cdot\bm{\hat{j}}_{\rm in}<0$), we can therefore expect that tides cause angular momentum to initially flow from the stellar rotation to the inner orbital motion and the eccentricity and semimajor axis to increase. Tides operate to circularise and contract the orbit only after the angular momentum flow peters out around $\bm{\Omega}_i\cdot\bm{\hat{j}}_{\rm in}/\omega_{\rm in}=f_2/f_5j_{\rm in}^3$ for which the r.h.s. of Eq.~\eqref{eq:L-Tide} becomes zero and of Eqs.~\eqref{eq:e-Tide} and~\eqref{eq:a-Tide} negative for any eccentricity value $0<e_{\rm in}<1$ \citep[][]{2011CeMDA.111..105C}. 

Furthermore, the torques exerted on the static tidal bulges induce a precession of $\bm{e}_{\rm in}$ about $\bm{\hat{j}}_{\rm in}$ on a timescale
\begin{equation}\label{eq:t-Tide}
    t_{\rm Tide}=1\Big/\sum_{i=1,2}15k_{\rm A}\omega_{\rm in}\frac{m_{(i-1)}}{m_i}\left(\frac{R_i}{a_{\rm in}}\right)^5\frac{f_4(e_{\rm in})}{j_{\rm in }^{10}},
\end{equation}
where $k_{\rm A}=0.014$ is the classical apsidal motion constant \citep{2007ApJ...669.1298F} and which is usually much shorter than the time by which tides could circularise the orbit.

The tidal description outlined above is more appropriate for stars with deep convective envelopes. Following \citet[][]{2002MNRAS.329..897H}, we include in {\tt TSE} a different tidal mechanism for stars which have a radiative envelope. In this case, the dominant tidal forces are dynamical and emerge from stellar oscillations which are excited by the binary companion \citep[][]{1975A&A....41..329Z,1977A&A....57..383Z}. In that case, we parametrise the tidal strength  by the lag time \citep[][]{1981A&A....99..126H}
\begin{equation}
    \tau_{1(2)}=\frac{R_{1(2)}}{Gm_{1(2)}T_{1(2)}},
\end{equation}
where 
\begin{align}
    \frac{k_A}{T_{1(2)}}=&1.9782\times10^4
    \left(\frac{m_{1(2)}}{\msun}\right)
    \left(\frac{R_{1(2)}}{\rm R_\odot}\right)^2
    \left(\frac{\rm R_\odot}{a_{\rm in}}\right)^5\nonumber\\
    &\times\left(1+\frac{m_{2(1)}}{m_{1(2)}}\right)^{5/6}\frac{E_{2,1(2)}}{\rm yr}
\end{align}
and
\begin{equation}
    E_{2,1(2)}\simeq 10^{-9}\left(\frac{m{_{1(2)}}}{\msun}\right)^{2.84}.
\end{equation}
Following \citet[][]{2002MNRAS.329..897H} the code applies dynamical tides for all MS stars with a mass greater than $1.25\,\msun$, core Helium Burning stars, and naked helium MS stars.

The coefficient $E_2$ is related to the structure
of the star and refers to the coupling between the tidal
potential and gravity mode oscillations. Its value is difficult to estimate since it is very
sensitive to the structure of the star and therefore to the
exact treatment of stellar evolution. Importantly, the equations of motion (\ref{eq:t-Tide})  were developed in \cite{1981A&A....99..126H} 
under the assumption that the tides reach an equilibrium shape with a constant time lag.
These equations hold for very small deviations
in position and amplitude with respect to the equipotential surfaces. Thus,
we caution that dynamical tides, where the stars oscillates radially,  are not properly described by the constant time-lag model. At every periastron passage, tidal stretching and compression
can force the star to oscillate in a variety of eigenmodes. The excitation and damping of these eigenmodes can significantly affect  the secular evolution of a binary orbit \citep[][]{2018AJ....155..118W,2019PhRvD.100f3001V,2019MNRAS.484.5645V}.

Because the physics of stellar tides is much uncertain
and the efficiency of tides itself is debated,
in the simulations presented here we consider two choices.
In our fiducial models
we opt for a simplified
approach in which we employ the equilibrium tide equations for all stars with a constant $\tau=1\,{\rm s}$, thus encapsulating all the uncertainties related to tides in this constant factor. In another set of models ({\tt Incl. dyn. tides}), we follow the approach of \citet[][]{2002MNRAS.329..897H} and use either equilibrium or dynamical tides depending on the stellar mass and type as described above. 

We find that our main results are not significantly affected by the implementation of dynamical tides in the code. In Section~\ref{sec:results}, we will therefore primarily focus on our fiducial choice with constant $\tau=1\,\rm s$.

\subsubsection{Rotational distortion (Rot)}
The rotation of each star distorts its shape inducing a quadrupole moment. As a result, the binary stars torque each other yielding \citep[e.g.,][]{2001ApJ...562.1012E}
\begin{align}
    \left.\frac{\dd\bm{e}_{\rm in}}{\dd t}\right\vert_\text{Rot}=&\sum_{i=1,2}\frac{k_{\rm A}m_{(i-1)}R_i^5}{2\omega_{\rm in}\mu_{12}a_{\rm in}^5}\frac{e_{\rm in}}{j_{\rm in}^4}\Bigg\{\Bigg[2\left(\bm{\Omega}_{i}\cdot\bm{\hat{j}}_{\rm in}\right)^2\nonumber\\
    &-\left(\bm{\Omega}_{i}\cdot\bm{\hat{q}}_{\rm in}\right)^2-\left(\bm{\Omega}_{i}\cdot\bm{\hat{e}}_{\rm in}\right)^2\Bigg]\bm{\hat{q}}_{\rm in}\nonumber\\
    &+2\left(\bm{\Omega}_{i}\cdot\bm{\hat{q}}_{\rm in}\right)\left(\bm{\Omega}_{i}\cdot\bm{\hat{j}}_{\rm in}\right)\bm{\hat{j}}_{\rm in}\Bigg\},\label{eq:e_A-Rot}\\
    \left.\frac{\dd\bm{S}_{\rm 1(2)}}{\dd t}\right\vert_\text{Rot}=&\sum_{i=1,2}\frac{k_{\rm A}m_{(i-1)}R_{i}^5}{\omega_{\rm in}\mu_{12}a_{\rm in}^5}\frac{L_{\rm in}}{j_{\rm in}^4}\left(\bm{\Omega}_{i}\cdot\bm{\hat{j}}_{\rm in}\right)\nonumber\\
    &\times\left[\left(\bm{\Omega}_{i}\cdot\bm{\hat{q}}_{\rm in}\right)\bm{\hat{e}}_{\rm in}-\left(\bm{\Omega}_{i}\cdot\bm{\hat{e}}_{\rm in}\right)\bm{\hat{q}}_{\rm in}\right],\\
    \left.\frac{\dd\bm{j}_{\rm in}}{\dd t}\right\vert_\text{Rot}=&-\frac{j_{\rm in}}{L_{\rm in}}\sum_{i=1,2}\left.\frac{\dd\bm{S}_{\rm i}}{\dd t}\right\vert_\text{Rot}.
\end{align}
Analogously to the tidal torques, the first term in the bracket of Eq.~\eqref{eq:e_A-Rot} causes the inner orbit's periapsis to precess about $\bm{\hat{j}}_{\rm in}$ on a timescale
\begin{equation}\label{eq:t-Rot}
    t_{\rm Rot}=1\Big/\sum_{i=1,2}\frac{k_{\rm A}m_{(i-1)}R_i^5}{2\omega_{\rm in}\mu_{12}a_{\rm in}^5j_{\rm in}^4}.
\end{equation}

\subsubsection{Mass-loss (Mass)}
During its lifetime, the mass of a star can substantially decrease as a result of e.g., stellar winds \citep{2000MNRAS.315..543H} and the explosive mass-loss in a SN \citep{1961BAN....15..265B}. If the mass-loss of the star is isotropic its spin simply changes as [cf., Eq.~\eqref{eq:S}]
\begin{equation}
    \left.\frac{\d \bm{S}_{1(2)}}{\d t}\right\vert_\text{Mass}=\bm{S}_{1(2)}\frac{\dot{m}_{1(2)}}{m_{1(2)}},
\end{equation}
where $\dot{m}_{1(2)}=\d m_{1(2)}/\d t \leq0$ is the mass-loss rate. While the stars lose mass, the specific orbital angular momentum $L_{\rm in}/\mu_{12}$ is conserved. Hence, the semimajor axis of the inner orbit changes as [cf., Eq.~\eqref{eq:L_in}]
\begin{equation}\label{eq:a_in-Mass}
    \left.\frac{\d a_{\rm in}}{\d t}\right\vert_\text{Mass}=-a_{\rm in}\frac{\dot{m}_{12}}{m_{12}},
\end{equation}
where $\dot{m}_{12}=\dot{m}_{1}+\dot{m}_{2}$, i.e. mass-loss loosens the binary since $\dot{m}_{12}<0$ implies $\d a_{\rm in}/\d t>0$. 

\subsubsection{Schwarzschild and de Sitter precession (1PN)}
At first post-Newtonian order, relativistic effects cause the eccentricity vector $\bm{e}_{\rm in}$ of the inner orbit to precess about the orbital axis $\bm{\hat{j}}_{\rm in}$ as 
\begin{equation}\label{eq:e_A-1PN}
    \left.\frac{\dd\bm{e}_{\rm in}}{\dd t}\right\vert_\text{1PN}=\frac{e_{\rm in}}{t_{\rm 1PN}}\bm{\hat{q}}_{\rm in},
\end{equation}
where we defined the associated timescale
\begin{equation}\label{eq:t-1PN}
    t_{\rm 1PN}=\frac{c^2a_{\rm in}^{5/2}j_{\rm in}^2}{3G^{3/2}m_{12}^{3/2}}.
\end{equation}
This apsidal precession is referred to as \citet{1916AbhKP1916..189S} precession. Also at first post-Newtonian order, we have the de Sitter precession of the stellar spins $\bm{S}_{1(2)}$ that are parallel-transported along the orbit
\begin{equation}
    \left.\frac{\dd\bm{S}_{1(2)}}{\dd t}\right\vert_\text{1PN}=\frac{S_{1(2)}}{t_{\bm{S}_{1(2)}}}\bm{\hat{j}}_{\rm in}\times\bm{\hat{S}}_{1(2)},
\end{equation}
where
\begin{equation}
    t_{\bm{S}_{1(2)}}=\frac{c^2a_{\rm in}j_{\rm in}^2}{2G\mu_{12}\omega_{\rm in}}\left[1+\frac{3m_{2(1)}}{4m_{1(2)}}\right]^{-1}.
\end{equation}

\subsubsection{Lense-Thirring precession (1.5PN)}
At 1.5 post-Newtonian order, the spins of the inner binary members back-react on the orbit inducing a frame-dragging effect. As a result, the orbit changes as \citep{PhysRevD.12.329,2018ApJ...863....7R}
\begin{align}
    \left.\frac{\dd\bm{e}_{\rm in}}{\dd t}\right\vert_\text{1.5PN}=&\frac{2G}{c^2}\sum_{i=1,2}\frac{S_i e_{\rm in}}{a_{\rm in}^3j_{\rm in}^3}\left(1+\frac{3m_{(i-1)}}{4m_i}\right)\nonumber\\
    &\times\left[\bm{\hat{S}}_i-3(\bm{\hat{S}}_i\cdot\bm{\hat{j}}_{\rm in})\right]\bm{\hat{e}}_{\rm in}\label{eq:e_A-1.5PN},\\
    \left.\frac{\dd\bm{j}_{\rm in}}{\dd t}\right\vert_\text{1.5PN}=&\frac{2G}{c^2}\sum_{i=1,2}\frac{S_i}{a_{\rm in}^3j_{\rm in}^2}\left(1+\frac{3m_{(i-1)}}{4m_i}\right)\bm{\hat{S}}_{i}\times\bm{\hat{j}}_{\rm in}.
\end{align}
The precessional term on the r.h.s. of Eq.~\eqref{eq:e_A-1PN} is larger than that of Eq.~\eqref{eq:e_A-1.5PN} by a factor $\sim L_{12}/S_{1(2)}>1$ for the stellar systems we are interested in. Hence, we do not consider the timescale associated with Eq.~\eqref{eq:e_A-1.5PN} in the criterion~\eqref{eq:timescales}. 

\subsubsection{Gravitational waves (GW)}
The gravitational waves emitted by the stars on the inner orbit carry away orbital energy and angular momentum. This drainage causes the semimajor axis and eccentricity of the inner orbit to shrink, i.e. to tighten and to circularise, respectively, as \citep{1964PhRv..136.1224P}
\begin{align}
    \left.\frac{\dd a_{\rm in}}{\dd t}\right\vert_\text{GW}&=-\frac{64}{5}\frac{G^3\mu_{12}m_{12}^2}{c^5a_{\rm in}^3(1-e_{\rm in}^2)^{7/2}}\left(1+\frac{73}{24}e_{\rm in}^2+\frac{37}{96}e_{\rm in}^4\right),\label{eq:a_A-GW}\\
    \left.\frac{\dd e_{\rm in}}{\dd t}\right\vert_\text{GW}&=-\frac{304}{15}\frac{G^3\mu_{12}m_{12}^2e_{\rm in}}{c^5a_{\rm in}^4(1-e_{\rm in}^2)^{5/2}}\left(1+\frac{121}{304}e_{\rm in}^2\right)\label{eq:e_A-GW}.
\end{align}
If gravitational wave emission were the only effect acting on the inner binary, in particular if the tertiary perturbation is negligible, its time to coalescence $\tau_{\rm coal}$ can be found by integrating Eqs.~\eqref{eq:a_A-GW} and \eqref{eq:e_A-GW} which yields 
\begin{equation}
    \tau_{\rm coal}=3.211\times10^{17}\,{\rm yr}\left(\frac{a_{\rm in}}{\AU}\right)^4\left(\frac{\msun}{m_{12}}\right)^2\left(\frac{\msun}{\mu_{12}}\right)F(e_{\rm in})\label{eq:t-GW},
\end{equation}
where 
\begin{equation}
    F(e_{\rm in})=\frac{48}{19}\frac{1}{g^4(e_{\rm in})}\int_0^{e_{\rm in}}\frac{g^4(e^\prime_{\rm in})j_{\rm in}^5(e^\prime_{\rm in})}{e^\prime_{\rm in}(1+\frac{121}{304}e_{\rm in}^{\prime2})}\dd{}e^\prime_{\rm in}
\end{equation}
can be evaluated numerically using $g(e_{\rm in})=e_{\rm in}^{12/19}j_{\rm in}^{-2}(e_{\rm in})(1+121e_{\rm in}^2/304)^{870/229}$. For instance, an equal-mass mass binary with $m_1=m_2=30\msun$ would merge within $10\,\rm Gyr$ if it is closer than $a_{\rm in}\lesssim0.1\AU$. Meanwhile, the direction of the eccentricity vector and the orbital axis remain unchanged. Hence, this dissipation effect can be included in the vectorial Eqs.~\eqref{eq:j_A} and \eqref{eq:e_A} as
\begin{align}
    \left.\frac{\dd \bm{e}_{\rm in}}{\dd t}\right\vert_\text{GW}&=\left.\frac{\dd e_{\rm in}}{\dd t}\right\vert_\text{GW}\bm{e}_{\rm in},\\
    \left.\frac{\dd \bm{j}_{\rm in}}{\dd t}\right\vert_\text{GW}&=-\left.\frac{\dd e_{\rm in}}{\dd t}\right\vert_\text{GW}\frac{e_{\rm in}}{j_{\rm in}}\bm{\hat{j}}_{\rm in}.
\end{align}
The magnitude of the r.h.s. of Eq.~\eqref{eq:a_A-GW} strongly increases for small $a_{\rm in}$ and $1-e_{\rm in}^2$. This property has an important implication for binary stars that evolve to compact objects. The emission of gravitational waves promotes a merger thereof if and only if they move on a sufficiently tight or eccentric orbit. Thus, the large eccentricity excitations caused by the Lidov-Kozai effect can abet a coalescence of the inner binary members in a triple system \citep[e.g.,][]{2002ApJ...578..775B,2011ApJ...741...82T,2012ApJ...757...27A,2014ApJ...781...45A,2015ApJ...799..118P,2017ApJ...836...39S,2018MNRAS.480L..58A,2018ApJ...863....7R,2018MNRAS.481.4907G,2018ApJ...856..140H}.

\subsubsection{Outer orbit evolution}

For the evolution of the outer orbit we can safely neglect the relativistic effects and the torques emerging from the tides and stellar rotations since they are suppressed by the larger semimajor axis $a_{\rm out}$. The evolution is thus solely given by
\begin{align}
    \frac{\dd\bm{j}_{\rm out}}{\dd t}&=\left.\frac{\dd\bm{j}_{\rm out}}{\dd t}\right\vert_\text{LK,Quad}+\left.\frac{\dd\bm{j}_{\rm out}}{\dd t}\right\vert_\text{LK,Oct},\label{eq:j_out}\\
    \frac{\dd\bm{e}_{\rm out}}{\dd t}&=\left.\frac{\dd\bm{e}_{\rm out}}{\dd t}\right\vert_\text{LK,Quad}+\left.\frac{\dd\bm{e}_{\rm out}}{\dd t}\right\vert_\text{LK,Oct},\label{eq:e_out}\\
    \frac{\dd a_{\rm out}}{\dd t}&=\left.\frac{\dd a_{\rm out}}{\dd t}\right\vert_\text{Mass}\label{eq:a_out},
\end{align}
where the Lidov-Kozai terms are given by Eqs.~(17)~--~(20) of \citet[][]{2015MNRAS.447..747L} and the mass-loss term is, analogously to Eq.~\eqref{eq:a_in-Mass}, given by
\begin{equation}\label{eq:a_out-Mass}
    \left.\frac{\d a_{\rm out}}{\d t}\right\vert_\text{Mass}=-a_{\rm out}\frac{\dot{m}_{123}}{m_{123}},
\end{equation}
Thus, we also do not follow the spin evolution of the outer companion.
Together, Eqs.~\eqref{eq:j_A}~--~\eqref{eq:S_1(2)} and \eqref{eq:j_out}~--~\eqref{eq:a_out} constitute a coupled set of twenty differential equations (vectorial quantities counting thrice) which
we integrate forward in time. Simultaneously, we keep track of the evolution of the stellar masses and radii, $m_{1(2)(3)}=m_{1(2)(3)}(t)$ and $R_{1(2)}=R_{1(2)}(t)$, respectively. This is governed by the rich stellar physics describing the coevolution of the three massive stars that we implement as discussed in the following section.

\subsection{Stellar evolution}\label{sec:evolution}
In the following, we describe our treatment of stellar evolution. The stars are evolved using the stellar evolution code  {\tt Single Stellar Evolution} \citep[{\tt SSE},][]{2000MNRAS.315..543H}. We modified this code to include up-to-date prescriptions for stellar winds, black hole formation, and SN kicks and we couple it to the equations above to account for the dynamical evolution of the system. 

We use metallicity-dependent stellar wind prescriptions  \citep{2001A&A...369..574V}. 
These are the same stellar evolution subroutines currently employed in other population synthesis codes \citep[e.g.,][]{2016ApJ...819..108B,2020ApJ...898...71B}. With these modifications, {\tt TSE} reproduces
the mass distribution for single black holes (BHs) adopted in recent studies of
compact object binary formation from field binaries and clusters \citep[e.g.,][]{2020A&A...636A.104B,2016PhRvD..93h4029R,2020A&A...639A..41B,2020arXiv200901861A}. Optionally, {\tt TSE} takes a mass-loss dependency on the electron-scattering Eddington factor into account \citep[][]{2008A&A...482..945G,2011A&A...535A..56G,2011A&A...531A.132V,2017RSPTA.37560269V,2018MNRAS.474.2959G}.

In {\tt TSE}, the initial radius of each star is given by {\tt SSE} where it is calculated from the initial mass and metallicity as in \citet{2000MNRAS.315..543H}.
By default, the initial spin  for each star is taken also to be consistent with the adopted value in {\tt SSE} where the equatorial speed of zero age MS stars is set equal to \citep{1992adps.book.....L}
\begin{equation}
v_{{\rm rot},1(2)}=330{\rm km\ s^{-1}}\left( {m_{1(2)}\over M_\odot}\right)^{3.3}\left[15+\left({m_{1(2)}\over M_\odot}\right)^{3.45} \right]^{-1} \ ,
\end{equation}
so that the initial spin frequency becomes $\Omega_i=v_{{\rm rot},1(2)}/R_{1(2)}$. For this work, the spins are assumed to be initially aligned with the orbital angular momentum of the binary.

When a star evolves to become a neutron star (NS) or a BH, the remnant radius is set to zero, and
its mass is immediately updated. In {\tt TSE}, the model adopted for the remnant masses is set by the code parameter {\tt nsflag}. If ${\tt nsflag}=1$ the BH and NS masses are computed as in \citet{2002ApJ...572..407B};
 if ${\tt nsflag}=2$ the BH and NS masses are computed as in \citet{2008ApJS..174..223B}; if
${\tt nsflag}=3$ they are given by the ``rapid'' SN prescription described in \citet{2012ApJ...749...91F}; and if ${\tt nsflag}=4$ they are described by the  ``delayed'' SN prescription  also from \citet{2012ApJ...749...91F}.

Given the large uncertainties in the natal kick velocities of BHs, we adopt three different models for their distributions. We assume that kick velocities are randomly oriented, then the assumed model for the BH kick velocity magnitude is set by the code parameter {\tt bhflag}.
If ${\tt bhflag}=0$ the natal kicks of all BHs and NSs are set to zero. In any other case we assume that NS kicks follow a Maxwellian distribution with dispersion $\sigma=265\rm km\ s^{-1}$ \citep{2005MNRAS.360..974H}. If ${\tt bhflag}=1$, the BHs receive the same momentum kick as NSs, i.e., the BH kick velocities are lowered by the ratio of NS mass (set to $1.5\msun$) to BH mass. We will refer to them as "proportional" kicks. If ${\tt bhflag}=2$ we assume that the BH kicks are lowered by the mass that falls back into the compact object according to
\begin{equation}\label{eq:fallback}
v_{\rm k}=v_{\rm k, natal}(1-f_{\rm fb}),
\end{equation}
where $f_{\rm fb}$ is the fraction of the ejected SN mass
that falls back onto the newly formed proto-compact
object, which is given by the assumed SN mechanism set by the parameter {\tt nsflag}.

What we are interested in is the change to the orbital elements due to
the mass-loss and birth kicks as the stars evolve towards their final states.
When a remnant is formed, we extract the velocity of the
natal kick from the adopted prescription. The kick is then self-consistently applied to the orbital elements of the system following \citet{2012MNRAS.424.2914P}.
Briefly, we draw a random phase from the mean anomaly and then apply the instantaneous kick, $\bm{v}_{\rm k}$, to
the initial velocity vector of that
component, $\bm{v}_0$. Thus, the new
angular momentum and eccentricity vectors (using the new orbital velocity vector and the same orbital position vector) are given by
\begin{align}
\bm{j}_{\rm new}={r\times \bm{v}_{\rm new}\over \sqrt{m_{12,\rm new}a_{\rm new}}}
\end{align}
and 
\begin{equation}
\bm{e}_{\rm new}={1\over Gm_{12,\rm new}}\left(\bm{v}_{\rm new}\times \bm{j}_{\rm new} \right)-{\bm{r}\over r},
\end{equation}
where $m_{12,\rm new}$  is the new total mass of the binary and $\bm{v}_{\rm new}=\bm{v}_0+\bm{v}_{\rm k}$.
The new semimajor axis is
\begin{equation}
a_{\rm new}=\left({2\over r}- {{v}_{\rm new}^2\over G m_{12,\rm new}} \right)^{-1}.
\end{equation}
If the kicks occur in one of the inner binary components,
we must also take care of the kick imparted on the centre of mass of the binary.
Thus, the change in the centre of mass velocity of the inner
binary is explicitly calculated. This change is then
added to the velocity arising from the BH natal kick,
and applied as $\bm{v}_{\rm new}$ to the outer binary \citep[e.g.,][]{2019MNRAS.484.1506L}.
As a result, the orientation of the orbital plane changes. Meanwhile, it is uncertain if the spin orientation of the compact remnants changes as well. For young pulsars, \citet[][]{2012MNRAS.423.2736N,2013MNRAS.430.2281N} found evidence that the spins align with their proper motion which could be explained by NS natal kicks defining a preferred direction for the subsequent angular momentum accretion of fallback material \citep[][]{2022ApJ...926....9J}. Thus, the spin-kick correlation is expected to be stronger for higher natal kicks. 
Here we adopt the assumption made in the literature that  natal kicks leave the spin orientations unchanged    \citep[e.g.,][]{2012MNRAS.424.2914P,2016ApJ...832L...2R,2018ApJ...863....7R,2019MNRAS.484.1506L}.

\subsection{Mass transfer}
If a star is bound to a close companion, it can experience a set of binary interactions, including accretion of mass.
Accretion onto a companion star can occur  
during either Roche lobe overflow
or when material is accreted from a stellar wind. We describe below our simplified treatment of these two possible modes of accretion.

\subsubsection{Wind accretion}
The material ejected as a wind can be partly accreted by the companion star, or self-accreted by the donor star itself. 
 Because of gravitational focusing, the accretion cross
section is generally much larger than the geometric cross section of the accretor and it is often expressed by the Bondi-Hoyle accretion radius  \citep{1944MNRAS.104..273B}
\begin{align}
R_{\rm acc}={2Gm_{\rm acc}\over v^2}
\end{align}
with $m_{\rm acc}$ the accretor mass and $v$ the relative velocity between the wind and the accretor star.
For a mass-loss rate $\dot{m}_{\rm wind}$ and a spherically symmetric wind, the accretion rate is given by 
\begin{align}\label{wacc}
\dot{m}_{\rm acc}=-\dot{m}_{\rm wind} 
\left( {m_{\rm acc}\over m_{\rm don}+m_{\rm acc}} \right)^2
\left(v_{\rm orb} \over v_{\rm wind} \right)^4
\end{align}
where $m_{\rm don}$ is the donor mass, $v_{\rm orb}$ is the orbital velocity, and $v_{\rm wind}$ is the wind velocity.

The accretion process will affect the mass and spin of the stars, as well as the orbital parameters of the triple, e.g.,  equations~\eqref{eq:a_in-Mass} and ~\eqref{eq:a_out-Mass}. However,  
its  formulation presents a number of difficulties. 
First, when the wind mass losing star is the tertiary we should take into account that accretion occurs onto a binary rather than a single object, and there is no simple prescription to  describe this \citep[e.g.,][]{2019ApJ...884...22A}. 
Moreover, there are major uncertainties in modeling the evolution of the binary orbit and stellar spins due to wind accretion, which would require careful geometrical considerations of how the mass flow is ultimately accreted onto the star surface \citep{1998ApJ...497..303M,2009ApJ...700.1148D,2013ApJ...764..169P}.
Fortunately, massive stars are characterised
by high wind velocities, typically a few thousand $\rm km\ s^{-1}$ \citep{1992ASPC...22..167P,2001ASSL..264..215C}. Moreover, both the inner and outer orbit of the progenitors of compact object triples tend to be relatively wide -- in order to avoid a merger of the inner binary  during an episode of unstable mass-transfer and to guarantee  dynamical stability.
 Thus, the last factor in Eq.~\eqref{wacc}  generally makes the accretion rate  several orders of magnitude smaller than the mass-loss rate. 
 Since in the systems we consider, wind-accretion tends to be of secondary importance and much less important than
accretion by atmospheric Roche lobe overflow,
we proceed in what follows 
with the assumption that changes in mass and angular momentum from material gained by a wind can be ignored, i.e.,
  we set $\dot{m}_{\rm acc}=0 $.
We redirect the reader to \citet{2020arXiv201104513H} for an approximate treatment of wind accretion in triples and higher multiplicity systems.

\subsubsection{Roche lobe overflow}
 If one of the stars
in the inner binary overflows its Roche lobe, matter  can  move through
the first Lagrangian point and be accreted by the companion star.
We assume that Roche lobe overflow begins when the stellar radius of an inner binary component satisfies
\begin{equation}
    {R_{1(2)}}>\frac{0.49\left[m_{1(2)}/m_{2(1)}\right]^{2/3}a_{\rm in}(1-e_{\rm in})}{0.6\left[m_{1(2)}/m_{2(1)}\right]^{2/3}+\ln\left\{1+\left[m_{1(2)}/m_{2(1)}\right]^{1/3}\right\}}.\label{eq:Roche-in}
\end{equation}
The theory of Roche love overflow is based on two stars in a circular
orbit in which complete corotation has been achieved \citep{1983ApJ...268..368E}. The modelling  of  mass-transfer in eccentric orbits is the subject of ongoing research \citep{2016ApJ...825...71D,2019ApJ...872..119H}, but remains 
an elusive subject overall. 
For want of a more detailed treatment, when condition~\eqref{eq:Roche-in} is met we evolve the binary using  the  binary  stellar  evolution analogous to {\tt SSE}, the code {\tt Binary Stellar Evolution} \citep[{\tt BSE},][]{2002MNRAS.329..897H}. Here, the binary is subject to instant synchronisation and circularises on the tidal friction timescale.
The various parameters that enter in the equations of motion of the binary (e.g., $K$, $k_{\rm A}$, $\tau$) are chosen to be consistent with those used in Eqs.~\eqref{eq:j_A}~--~\eqref{eq:S_1(2)}.
During the entire episode of mass transfer we neglect the dynamical influence of the tertiary. 

Although necessarily approximate,  our approach is in most cases adequate because tides generally act on a time-scale shorter than the secular evolution time-scale of the triple, quenching the dynamical influence of the tertiary star. For example, using Eq.~\eqref{eq:timescales} it is easy to show that for equal mass components, the precession of the inner binary periapsis due to tidal bulges will fully quench the 
Lidov-Kozai oscillations for any $a_{\rm out}j_{\rm out}/a_{\rm in}\gtrsim 10j_{\rm in}^{3}/\left[f_4(1-e_{\rm in})^{5}\right]^{1/3}$, where $f_4=f_4(e_{\rm in})$ is a polynomial given in Appendix~\ref{sec:Hut}.
Moreover, when mass transfer begins at high eccentricities, dissipative tides can become dominant very quickly, circularising the orbit and thereby reducing the dynamical effect of the tertiary.

Finally, we assume that the tertiary star overfills its Roche lobe when  
\begin{equation}
    R_3>\frac{0.49{q_{\rm out}}^{2/3}}{0.6{q_{\rm out}}^{2/3}+\ln\left\{1+{q_{\rm out}}^{1/3}\right\}}a_{\rm out}(1-e_{\rm out})\label{eq:Roche-out}.
\end{equation}
Currently, we do not try to model mass transfer from the tertiary to the inner binary. Thus, if the previous condition is satisfied, we simply stop the integration.
\begin{figure*}
\vspace{-5pt}
\begin{multicols}{3}
    \includegraphics[height=1.15\linewidth]{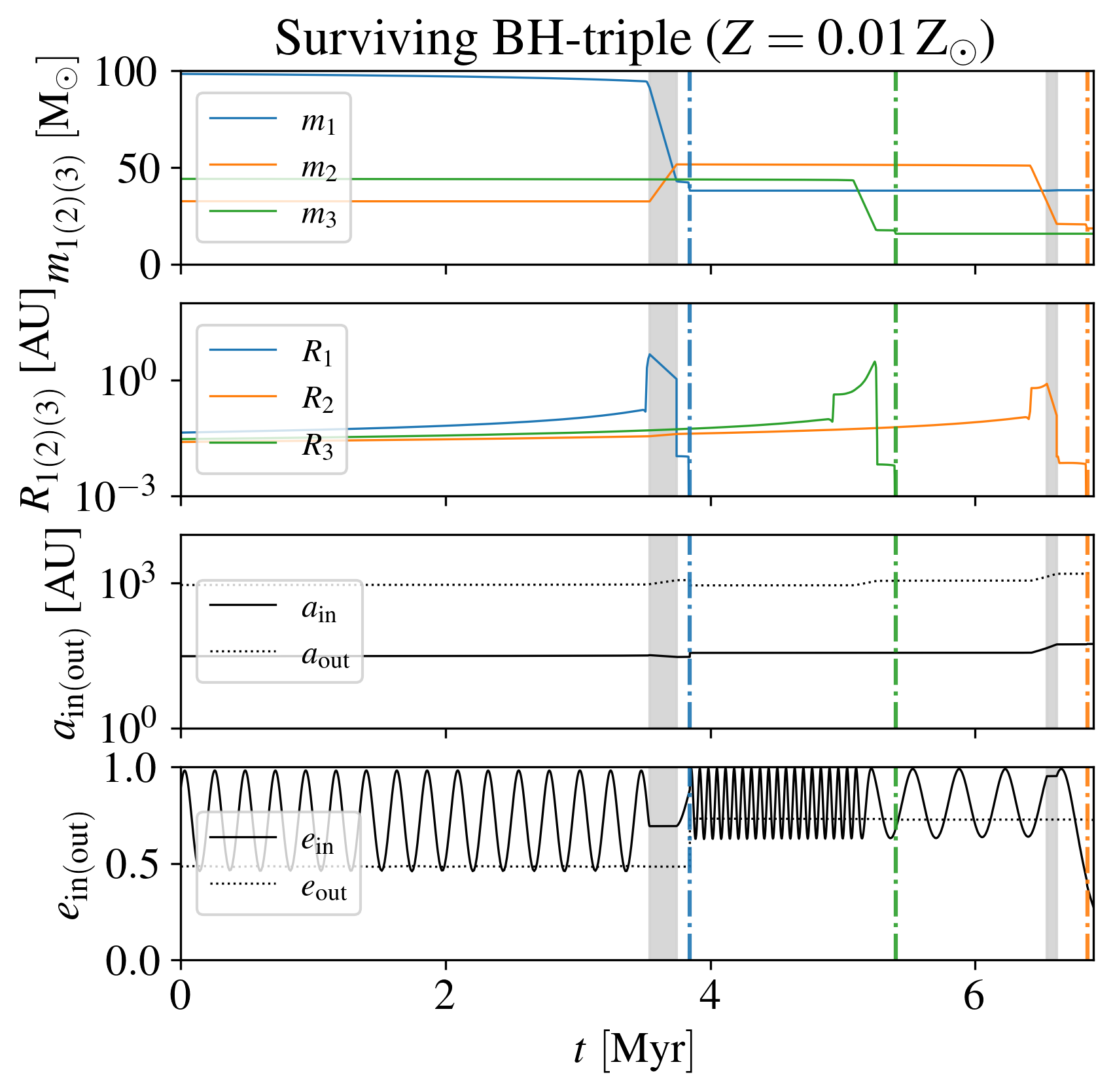}\par
    \includegraphics[height=1.15\linewidth]{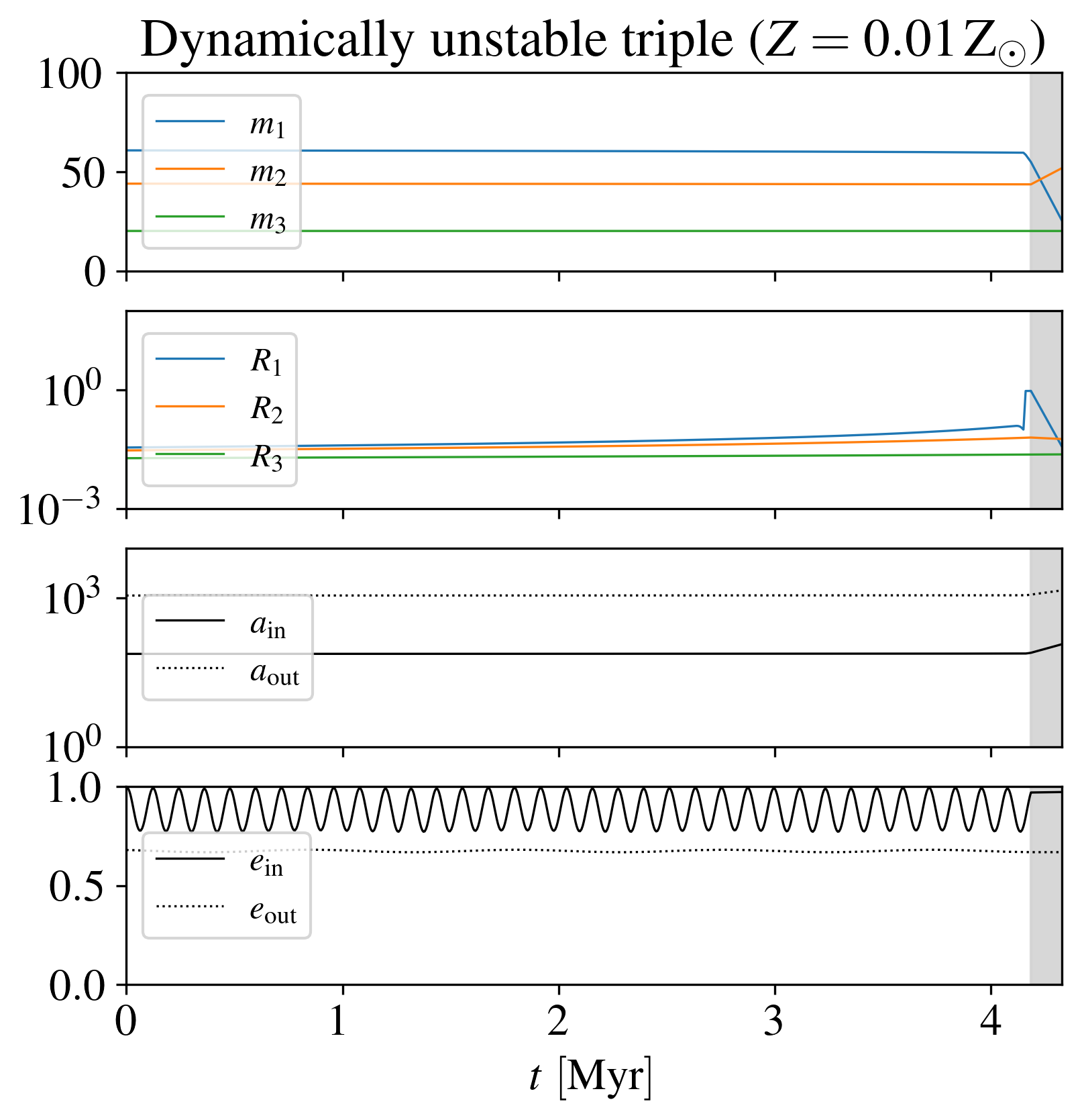}\par
    \includegraphics[height=1.15\linewidth]{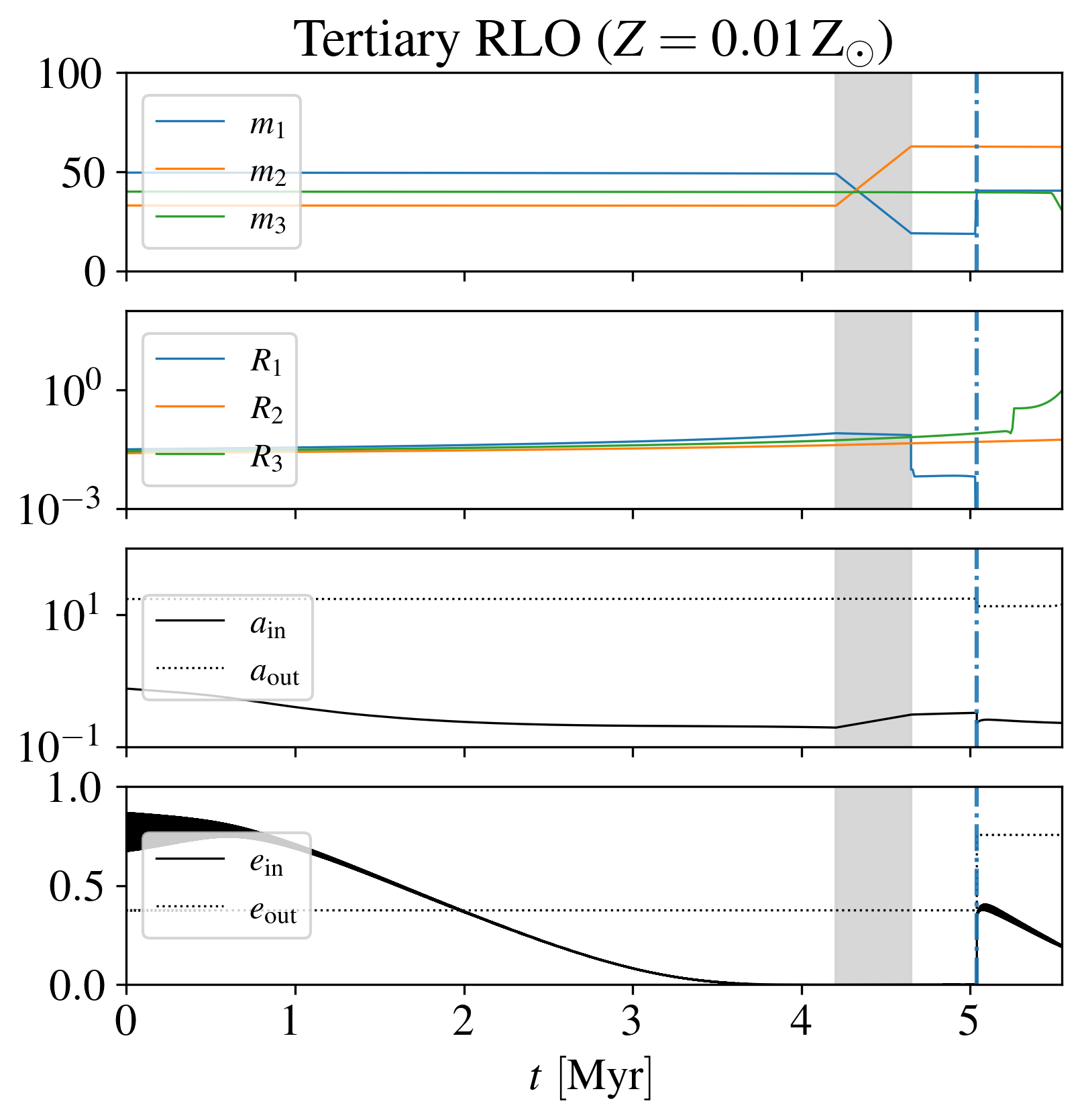}\par
\end{multicols}
\caption{Examples of the evolution of three stellar triples. Vertical dashed lines and grey shaded regions indicate the time of compact object formation and episodes of mass transfer in the inner binary, respectively. The initial parameters of the three triples are given in Appendix~\ref{appendix:examples}.}
\vspace{-5pt}
\label{fig:examples}
\end{figure*}

\subsection{Coupling stellar evolution and dynamics}

In the code presented in this paper, stellar evolution and dynamics  are coupled by using the following numerical treatment.

Because we neglect wind mass accretion, the mass and radius of each star will evolve as if they were isolated, at least until the next Roche lobe overflow episode occurs.
Thus, we start by setting a final integration time and compute the evolution of the stellar masses and radii using {\tt SSE} until this final time is reached.
Simultaneously, we use these masses and radii as a function of time in Eqs.~\eqref{eq:j_A}~--~\eqref{eq:S_1(2)} to determine the evolution of the stellar orbits and spins. During the integration of the equations of motion we check whether any of the stars forms a compact object. If they do, we calculate the natal kick according to the adopted prescriptions and compute the effect of the kick on the inner and outer orbits. 

Due to its lower binding energy the outer orbit is more vulnerable to disruptions than the inner one. As a consequence, there are some SN kicks which destroy the outer orbit while leaving the inner intact, i.e. the inner binary loses its tertiary companion. In this case, we continue the evolution of the remaining orbit with {\tt BSE}.

During the evolution, we check whether the system undergoes a phase of Roche lobe overflow. If mass transfer does not occur at any point during the evolution, the dynamical equations of motion are simply integrated until the required final time is reached. 

If a phase of Roche lobe overflow occurs in the outer binary, we stop the simulation.
If the mass-transfer phase occurs in the inner binary instead, we pass the required stellar and orbital parameters to {\tt BSE}  and continue evolving the binary until the end of the mass-transfer phase. 
During the {\tt BSE} integration, appropriate prescriptions from \citet{2002MNRAS.329..897H} are used to identify whether the stars come into contact and coalesce, if the binary reaches a common-envelope (CE) state, or if the mass-transfer is stable. If a merger occurs, we terminate the simulation. In particular, we assume that any CE evolution that is initiated by a donor star in the Hertzsprung gap (HG) leads to a stellar merger because it is questionable whether they already developed a well-defined core-envelope structure \citep[][]{2007ApJ...662..504B}. In the absence of a stellar core no stable binary configuration could result from a CE evolution. 
If the binary survives the mass transfer phase, we keep evolving the two inner stars with {\tt SSE} from the end of the mass transfer phase until the final integration time, and obtain new $m_{1(2)}(t)$ and $R_{1(2)}(t)$.
In this latter case, we store the orbital and stellar parameters  at the time the mass-transfer phase terminates and integrate Eqs.~\eqref{eq:j_A}~--~\eqref{eq:S_1(2)} from that moment on,  but using  the newly computed   $m_{1(2)}(t)$ and $R_{1(2)}(t)$.
Note that the stellar spins, $S_{1(2)}$, at the end of the mass transfer phase are assumed to be synchronised with the orbit, which is consistent with the treatment in {\tt BSE}.
Moreover, during the {\tt BSE} integration we use Eq.~\eqref{eq:a_out-Mass} to keep track of the evolution of $a_{\rm out}$ due to  mass-loss from the system.

\subsubsection{Stopping conditions}
In summary, the simulation is terminated before the final integration time in one of the following events:
\begin{enumerate}
    \item The tertiary star initiates a mass transfer episode onto the inner binary once it fills its Roche lobe according to Eq.~\eqref{eq:Roche-out}.
    \item The inner binary stars merge after an unstable mass transfer phase or an eccentric encounter. 
    \item The triple becomes dynamically unstable (see Section~\ref{sec:discarding}).
    \item The inner orbit is disrupted due to a SN.
\end{enumerate}
Either of these events leads to very different evolutionary outcomes. A tertiary RLO (i) may occur stably or initiate a CE engulfing all three stars in which a merger of two stars, chaotic ejection of one of them, or of the envelope is possible \citep[][]{2020MNRAS.491..495D,2020MNRAS.493.1855D,2021MNRAS.500.1921G,2021arXiv211000024H}. Yet, modelling tertiary RLO is less understood than RLO in isolated binaries due to the additional complexity of the inner binary motion.

If the inner binary merges before the formation of compact objects (ii),
a post-merger binary can form which consists of a massive post-merger star and the tertiary companion \citep[][]{2019Natur.574..211S,2020MNRAS.495.2796S,2021MNRAS.503.4276H}. If the initial triple was sufficiently compact a merging binary black hole might eventually from from the stellar post-merger binary \citep[][]{2022arXiv220316544S}.

Triples that become dynamically unstable (iii) can no longer be described by our secular approach, but enter a chaotic regime in which the ejection of one star or the merger of two become likely \citep[][]{2001MNRAS.321..398M,2015ApJ...808..120P,2022A&A...661A..61T}.

Lastly, if a SN disrupts the inner binary (iv), we expect that either the outer binary is also disrupted  due to the kick imparted to the inner binary centre of mass, or the remaining inner binary star and tertiary companion subsequently evolve on a wide orbit.

\begin{table}
	\centering
	\caption{Model parameters. In all models
	we also set 
	${\tt nsflag=3}$ (rapid SN prescription),
	$\alpha_{\rm CE}=1$, and
	$\tau=1$s.}	
	\label{tab:parameters}
	\begingroup
	\renewcommand*{\arraystretch}{1.2}
	\begin{tabular}{cccc}
	\hline
	\hline
	\multirow{2}{*}{Model} & {Metallicity Z} &  \multirow{2}{*}{\tt bhflag} & $\tau$\\
	& $[\rm Z_\odot]$ & &  [s] \\
    \hline
	\hline
    {\tt Fallback kicks} & 0.01, 1.0 & 2 & 1.0 \\
    {\tt Proportional kicks} & 0.01, 1.0 & 1 & 1.0 \\
    {\tt No kicks} & 0.01, 1.0 & 0 & 1.0 \\
    {\tt Incl. dyn. tides} & 0.01, 1.0 & 2 & See Sec.~\ref{sec:tides} \\
    \hline
	\hline
	\end{tabular}
	\endgroup
\end{table}

\subsection{Stellar evolution parameters} 
In this work, we investigate a set of different models whose parameters are summarised in Table~\ref{tab:parameters}. In any of our models we set the common-envelope efficiency parameter $\alpha_{\rm CE}$ to $1$ and the tidal lag time $\tau$ to $1\,\rm s$. The latter recovers well the observation of circularised inner binaries at short periods. The remnant masses prescription follows the "rapid" SN model \citep[${\tt nsflag=3}$,][]{2012ApJ...749...91F}. We study the impact of natal kicks by adopting the three models {\tt fallback kicks}, {\tt proportional kicks}, and {\tt no kicks} in which we set {\tt bhflag} to 2, 1, and 0, respectively, and investigate the effect of metallicity by setting $Z=0.01\zsun$ (low metallicity) or $Z=1.0\zsun$ (high/solar metallicity). If not stated differently, the {\tt fallback kicks} model is used as a default in the following sections.

\subsection{Example cases}
In Figure~\ref{fig:examples}, we show the evolution of three example systems at $Z=0.01\zsun$. The systems in the left and middle panels undergo LK oscillations, while in the right panel we see a system where the oscillations are quickly quenched by the tides acting between the  inner binary stars. All three systems enter one or two phases of stable mass transfer, which are indicated by the vertical grey shaded regions. 
As a consequence of the mass and semi-major axes changes, the period and maximum eccentricity of the LK oscillations in the system of the left panel changes after the mass transfer episode, which produces the observed modulation. A similar effect can be seen after the formation of a BH as indicated by the vertical dashed lines.

The system in the left panel survives all peculiar steps during the stellar evolution and eventually ends up as a stable BH triple. This is not the case for the system shown in the middle panel. Here, the expansion of the inner binary during a mass transfer phase causes the triple to become dynamically unstable (see Section~\ref{sec:discarding}). In contrast, the system in the right panel starts relatively  compact with an initial outer semi-major axis of only $a_{\rm out}\approx17.2\,\rm AU$. This is small enough for the tertiary companion to fill its Roche lobe during its giant phase. Then, we stop the integration for want of a more accurate treatment.

In Appendix~\ref{appendix:examples}, we list the initial parameters of the three exemplary triples. 

\section{Initial conditions}\label{sec:initial-conditions}
\begin{figure*}
\vspace{-5pt}
\centering
\includegraphics[width=0.7\textwidth]{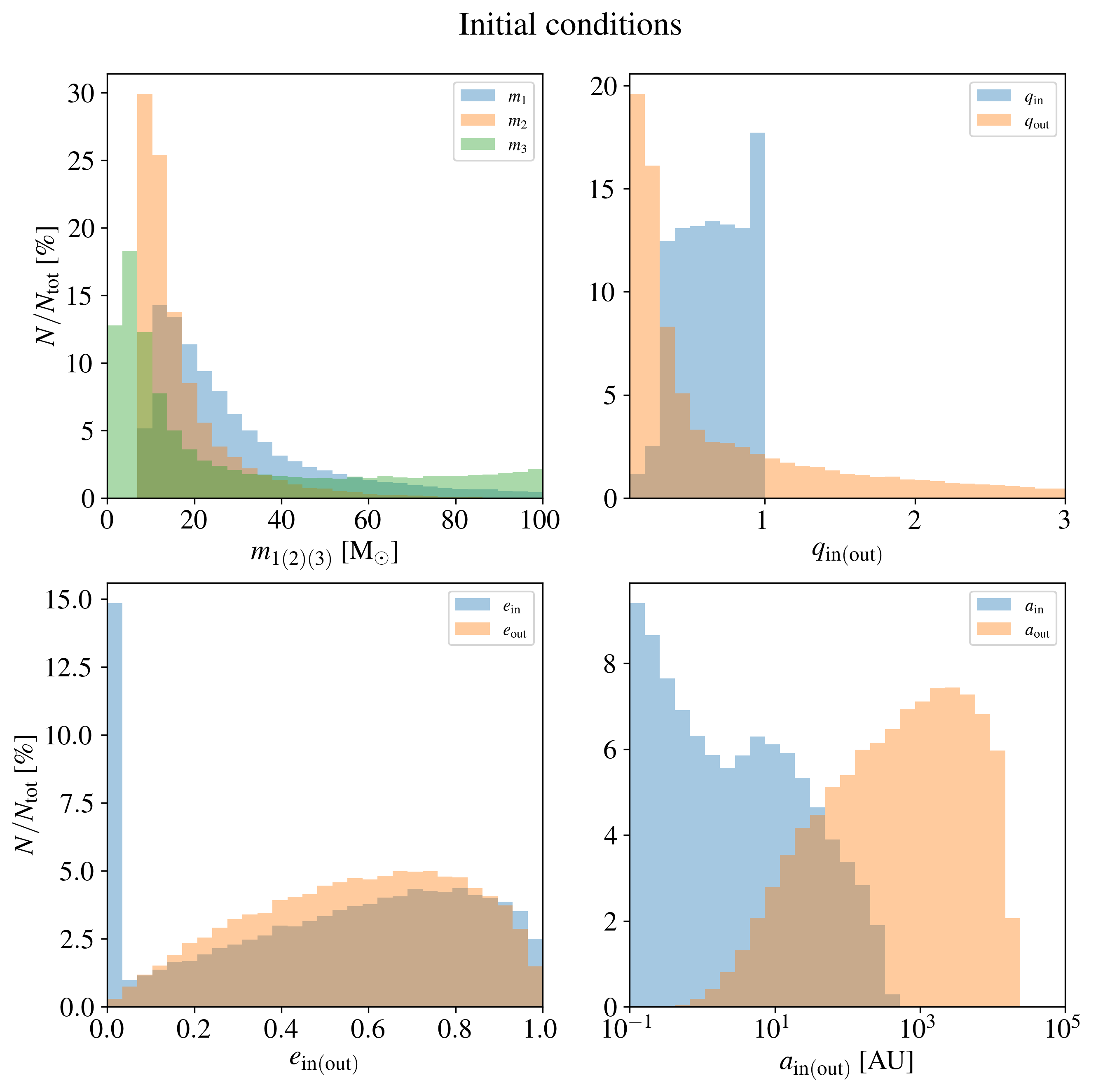}
\caption{Initial conditions of the triple population. The counts $N$ are normalised w.r.t. the total number of triples $N_{\rm tot}$ which are initially stable, detached, and whose inner binary members are massive enough to form compact objects.}
\vspace{-5pt}
\label{fig:MdS}
\end{figure*}

In the following, we describe the set-up of the initial parameter distribution of our massive stellar triple population. The initial time is chosen when the stars are on the ZAMS. Observationally, companions to massive, early-type stars were discovered by means of several techniques, e.g., radial velocity monitoring \citep[e.g.,][]{2001A&A...368..122G,2012Sci...337..444S,2014ApJS..213...34K}, eclipses \citep[e.g.,][]{2016AJ....151...68K,2016AcA....66..405S,2016AcA....66..421P,2015ApJ...810...61M}, proper motion \citep[e.g.,][]{2007AJ....133..889L}, and interferometry \citep[e.g.,][]{2013MNRAS.436.1694R,2014ApJS..215...15S}. For massive triples, it has been shown that the parameter distributions of early-type stars are a good indicator for the initial distribution at birth \citep[][]{2019MNRAS.488.2480R}. For the initial parameter distribution of our population we follow \citet[]{2017ApJS..230...15M} who compiled the variety of previous surveys. 

Accordingly, the masses and mass ratios, eccentricities, and orbital periods are not statistically independent from each other. Instead, they show important correlations across different periods, e.g. an excess of nearly-equal mass ratios ("twins") and circularised orbits at short periods whereas the properties of two stars are more consistent with a random pairing process toward long periods.

Specifically, we adopt the following sampling procedure which results in the marginalised distributions shown in Figure~\ref{fig:MdS}. At first, we propose an inner binary from the joint probability distribution
\begin{align}
    f(m_1,m_2,P_{\rm in},e_{\rm in})&=f(m_1)f(P_{\rm in}|m_1)\nonumber\\
    &\times f(m_2|m_1,P_{\rm in})f(e_{\rm in}|m_1,P_{\rm in}),\label{eq:two-third}.
\end{align}
Afterwards, an outer orbit is repeatedly drawn from the distribution
\begin{align}
    f(m_3,P_{\rm out},e_{\rm out}|m_1)&=f(P_{\rm out}|m_1)\nonumber\\
    &\times f(\tilde{q}_{\rm out}|m_1,P_{\rm out})f(e_{\rm out}|m_1,P_{\rm out}),
\end{align}
with $\tilde{q}_{\rm out}=m_3/m_1$, until the triple system is hierarchically stable and detached (see below). 

This procedure recovers the observed distributions of triples in which $m_1$ is the largest mass of the triple stars, i.e. it is part of the inner binary. Unfortunately, triples where the tertiary companion is the most massive star completely elude detection since it is difficult to resolve additional companions to the less massive star of a wide orbit. In order to model those kind of systems we agnostically draw in every third system the tertiary mass from a uniform distribution with a lower limit of $m_1$ and the orbital parameters from
\begin{align}
    f(P_{\rm out},e_{\rm out}|m_3)&=f(P_{\rm out}|m_3)f(e_{\rm out}|m_3,P_{\rm out}).
\end{align}

The triples proposed in this way are only retained if they are hierarchically stable and detached which naturally skews the inner and outer orbital distributions. The marginal distributions are as following \citep[]{2017ApJS..230...15M}. For convenience we define $m_{\rm p}={\rm max}(m_1,m_3)$ to be the largest mass of the triple. 

\subsection{Primary mass distribution $f(m_1)$}
The primary star is the more massive component of the inner binary. We draw its mass between $8$ and $100\msun$ from the canonical initial mass function \citep{2001MNRAS.322..231K} which is described by a single power law $f(m_1)\d m_1\propto m_1^\alpha\d m_1$ with exponent $\alpha=-2.3$. In general, the canonical initial mass function describes the mass distribution of all stars that formed together in one star-forming event. Note that it does not necessarily coincide with the initial mass distribution of the primaries which is skewed towards larger masses. However, for the \textit{massive} primaries under consideration both are approximately equal \citep[][Section 9]{2013pss5.book..115K}.

\subsection{Period distributions $f(P_{\rm in(out)}|m_{\rm p})$}  
The inner and outer periods $P_{\rm in(out)}$ are technically proposed from the same conditional distribution $f(P_{\rm in(out)}|m_{\rm p})$ in the range $0.2\leq\log_{10}(P_{\rm in(out)}/\rm day)\leq 5.5\,(8.0)$. This distribution function is slightly bimodal with one dominant peak at short periods, $\log_{10}(P_{\rm in(out)}/\rm day)<1$ \citep[consistent with][]{2012Sci...337..444S}, and another at $\log_{10}(P_{\rm in(out)}/\rm day)\approx3.5$. Discarding hierarchically unstable triples (see Section~\ref{sec:discarding}), roughly $41\%$ ($0\%$) of the systems have inner (outer) periods below $10\,\rm days$, $86\%$ ($10\%$) below $10^3\,\rm days$, and $99\%$ ($48\%$) per cent below $10^5\,\rm days$. After specifying the mass ratios (see below), the resulting semi-major axis distributions are shown in the lower right panel of Figure~\ref{fig:MdS}.

\subsection{Inner (outer) mass ratio distribution $f(q_{\rm in}(\tilde{q}_{\rm out})|m_{\rm p},P_{\rm in(out)})$}
The mass ratio distributions are described by an underlying broken power-law with two slopes $\alpha=\alpha_{\rm smallq}(m_{\rm p},P_{\rm in(out)})$ and $\alpha_{\rm largeq}(m_{\rm p},P_{\rm in(out)})$ for $0.1\leq q<0.3$ and $q\geq0.3$, respectively. This is shown in the upper right panel of Figure~\ref{fig:MdS}. Small inner mass ratios are further reduced since we only retain secondary stars with a mass $m_{\rm 2}\geq8\msun$. Moreover, observational surveys of massive primaries have discovered an excess fraction of twins \citep[][]{2000A&A...360..997T,2006ApJ...639L..67P}, i.e. companions with a mass similar to their primary ($q_{\rm in}>0.95$), if their orbital period is very short $\log_{10}(P_{\rm in}/\rm day)\lesssim1$, which gives rise to the large peak in the rightmost bin of the inner mass ratio distribution. In turn, the outer companion masses at long orbital periods are more consistent with a random pairing from the initial mass function \citep[][]{2017ApJS..230...15M}.

Since we are interested in inner binary stars which could form compact objects, their masses are restricted to $m_{1,2}\geq8\msun$. This restriction does not apply to the tertiary companion. Instead, we take any mass down to $m_3=0.1\msun$ into account.

\subsection{Inner (Outer) eccentricity $f(e_{\rm in(out)}|m_{\rm p},P_{\rm in(out)})$}
The inner (outer) eccentricity $e_{\rm in(out)}$ is drawn from the conditional distribution $f(e_{\rm in(out)}|m_{\rm p},P_{\rm in(out)})$ between $0$ and $1$. The distribution is fitted by an underlying power-law with exponent $\alpha=\alpha(P_{\rm in(out)})$ described as \citep[]{2017ApJS..230...15M}
\begin{equation}
    \alpha=0.9-\frac{0.2}{\log_{10}(P_{\rm in}/\rm day)-0.5}.
\end{equation}
In general, a power-law diverges at the lower boundary $e_{\rm in(out)}=0$ and cannot be interpreted as a probability density function if $\alpha\leq-1$. Here, this is the case if $\log_{10}(P_{\rm in}/\rm day)\lesssim0.6$. For these short periods it is reasonable to assume that all orbits were circularised due to tidal interactions \citep[e.g.,][]{1981A&A....99..126H,1989A&A...220..112Z,2001ApJ...562.1012E}. 

For longer periods, the power-law exponent increases monotonically where there is a narrow window, $0.6\lesssim\log_{10}(P_{\rm in}/\rm day)\lesssim0.7$, for which $-1<\alpha<0$ (i.e. the eccentricity distribution is skewed towards small values) and $\alpha\geq0$ for $\log_{10}(P_{\rm in}/\rm day)\gtrsim0.7$ (i.e. skewed towards large values). For long periods, the power-law approaches a thermal distribution. Note that \citet[]{2017ApJS..230...15M} imposed an approximate upper limit $e_{\rm max}(P_{\rm in(out)})<1$ for the eccentricity above which a binary is semi-detached or in contact at periapsis. Here, we explicitly check for each system whether one of the three stars initially fills its Roche lobe at periapsis and reject them as described below.

\subsection{Orbital angles}
We sample the initial values of the two arguments of periapsis of the inner and outer orbit and their relative inclination $i$ from isotropic distributions. The longitudes of the ascending nodes are "eliminated" by setting their difference to $\pi$ \citep[][]{2013MNRAS.431.2155N}.

Our assumption for the inclination distribution is uniformative since there exists no observational evidence about the mutual inclination $i$ for massive triples. Meanwhile, \citet[]{2016MNRAS.455.4136B} found all compact solar-type triples within $a_{\rm out}<10\AU$ have $i<60^\circ$, and the majority had $i<20^\circ$. Similarly, \citet[]{2017ApJ...844..103T} found nearly all triples with $a_{\rm out}<50\AU$ were prograde ($i<90^\circ$), and solar-type triples had random orientations only beyond $a_{\rm out}>10^3\AU$. However, he did note that more massive triples may be more misaligned, i.e., A/early-F triples achieved random orientations beyond $a_{\rm out}>100\AU$ (instead of $>10^3\AU$). If the overall preference of close solar-type triples for prograde inclinations turns out to persist in future observations of massive triples our isotropic assumption must be skewed towards small angles beyond the Kozai regime (cf. Section \ref{sec:EoM}).

\subsection{Discarded systems}\label{sec:discarding}
Triples that are proposed according to the sampling procedure described above are discarded if they are dynamically unstable, if at least one star fills its Roche lobe, or if the inner binary members are not massive enough to form compact objects ($m_{1(2)}<8\msun$; see \citet{2020A&A...640A..16T} for a study with less massive inner binaries). For the former two criteria we reject all systems that initially satisfy either 
\begin{align}\label{eq:stability}
    \frac{a_{\rm out}(1-e_{\rm out})}{a_{\rm in}}&<2.8\left[\left(1+\frac{m_3}{m_{12}}\right)\frac{1+e_{\rm out}}{\sqrt{1-e_{\rm out}}}\right]^{2/5},
\end{align}
or Eqs.~\eqref{eq:Roche-in} and~\eqref{eq:Roche-out}  \citep[]{2001MNRAS.321..398M,1983ApJ...268..368E,2019ApJ...872..119H}.

\subsection{Drawbacks in initial conditions}
Most previous population synthesis studies assume \mbox{(log-)uniform} initial distributions of the inner and outer mass ratios, orbital periods, semi-major axes, or eccentricities \citep[e.g.,][]{2017ApJ...841...77A,2017ApJ...836...39S,2018ApJ...863....7R,2020MNRAS.493.3920F,2021arXiv211000024H}. Typically, a mutual dependency of the orbital parameters is introduced by discarding initially unstable or Roche lobe filling systems, which, e.g., removes systems with relatively small inner semi-major axes and large inner eccentricities \citep{2017ApJ...841...77A,2020A&A...640A..16T}.
The drawback of  this procedure 
is that it fails at reproducing the known parameter distributions of  the inner binaries.
For example, consider a model in which the inner orbital periods are drawn from a given distribution that is inferred by observations \citep[e.g.,][]{2012Sci...337..444S}, whereas the outer semi-major axis distribution is uninformative (e.g., log-uniform), reflecting our poor statistics on wide (outer) binaries. 
A large number of triples will be discarded based on Eq.~\eqref{eq:stability} because they are dynamically unstable. As a consequence, the resulting orbital distribution of the inner binaries will deviate from the observationally motivated model that was assumed in the first place. Moreover, the adopted method does not take into account the observed correlation between the different orbital parameters of early-type stars.

The sampling procedure presented in this paper aims to improve previous work by reproducing some of the statistical features identified by observations \citep[][]{2017ApJS..230...15M}. Thus, the novel feature of our method is that it takes into account the observed mutual correlation between orbital parameters. Moreover, the distributions of the inner binary properties in our triple systems are consistent with observations since for a given inner binary we propose a tertiary until the triple satisfies the stability criteria.
But, we remain speculative regarding triples in which the most massive component is the tertiary star and for which there are no observations. Since the Lidov-Kozai effect is stronger for larger tertiary masses (cf. Section~\ref{sec:Lidov}), this introduces some uncertainty to the total fraction of systems in which a tertiary can dynamically perturb the inner binary.

\begin{table*}
	\centering
	\caption{Upper half: Fraction of triple evolutionary outcomes for our different models at sub-solar and solar metallicity. The last three columns refer to the fraction of surviving systems that harbour a BBH ($\Gamma_{\rm BBH}$), NSBH ($\Gamma_{\rm NSBH}$), and BNS ($\Gamma_{\rm BNS}$) in the inner binary. For those and for the stellar mergers we provide the fraction of systems that retain their tertiary companion plus ("+") the systems that lose it in a SN explosion, i.e. keep evolving as isolated inner binaries. Lower half: Evolutionary outcomes of isolated inner binaries when no tertiary companion is included from the beginning of the simulation.}
	\label{tab:table}
	\begingroup
	\renewcommand*{\arraystretch}{1.2}
	\begin{tabular}{ccccccccccc}
	\hline
	\hline
	\multirow{3}{*}{Z $[\rm Z_\odot]$} & \multirow{3}{*}{Model} & \multirow{3}{*}{$N_{\rm tot}$} & \multicolumn{7}{c}{Fraction of evolutionary outcomes $N/N_{\rm tot}$ [$\%$]}\\
	& & & Orbital & Stellar & Dynamically & Tertiary & \multirow{2}{*}{$\Gamma_{\rm BBH}$} & \multirow{2}{*}{$\Gamma_{\rm NSBH}$} & \multirow{2}{*}{$\Gamma_{\rm BNS}$}\\
	& & & disruption & merger & unstable & RLO & & & \\
    \hline
	\hline
    \multirow{4}{*}{0.01} & {\tt Fallback kicks} & 71936 & 49.72 & 18.70 + 9.90 & 7.15 & 4.86 & 3.56 + 5.89 & 0.05 + 0.15 & 0.00 + 0.02 \\
    & {\tt Proportional kicks} & 65858 & 53.93 & 15.04 + 13.25 & 7.21 & 4.83 & 0.29 + 5.31 & 0.02 + 0.13 & 0.00 + 0.00 \\
    & {\tt No kicks} & 42891 & 9.57 & 28.81 + 23.89 & 9.93 & 5.04 & 5.27 + 7.68 & 1.14 + 7.34 & 0.10 + 1.25 \\
    & {\tt Incl. dyn. tides} & 9746 & 49.39 & 19.75 + 8.93 & 7.81 & 4.77 & 3.53 + 5.60 & 0.03 + 0.18 & 0.00 + 0.00 \\
    \hline
    \multirow{4}{*}{1.0} & {\tt Fallback kicks} & 104643 & 57.92 & 17.19 + 9.09 & 9.45 & 5.52 & 0.26 + 0.54 & 0.00 + 0.03 & 0.00 + 0.00 \\
    & {\tt Proportional kicks} & 75607 & 59.64 & 15.77 + 9.96 & 9.05 & 5.53 & 0.00 + 0.04 & 0.00 + 0.00 & 0.00 + 0.00 \\
    & {\tt No kicks} & 59020 & 9.47 & 33.28 + 23.26 & 14.26 & 5.79 & 1.74 + 2.71 & 1.37 + 5.79 & 0.14 + 2.18 \\
    & {\tt Incl. dyn. tides} & 14973 & 55.77 & 19.66 + 8.06 & 10.50 & 5.20 & 0.29 + 0.51 & 0.00 + 0.02 & 0.01 + 0.00 \\
    \hline
    \hline
    \multirow{3}{*}{0.01} & {\tt Fallback kicks} & 49598 & 58.49 & 30.20 &  &  & 11.11 & 0.19 & 0.02 \\
    & {\tt Proportional kicks} & 49614 & 63.24 & 29.59 & & & 7.00 & 0.15 & 0.01 \\
    & {\tt No kicks} & 49705 & 11.33 & 62.88 & & & 14.89 & 9.34 & 1.56 \\
    \hline
    \multirow{3}{*}{1.0} & {\tt Fallback kicks} & 47848 & 67.86 & 31.40 & & & 0.73 & 0.00 & 0.00 \\
    & {\tt Proportional kicks} & 47883 & 69.83 & 30.16 & & & 0.00 & 0.00 & 0.00 \\
    & {\tt No kicks} & 47789 & 9.75 & 74.35 & & & 4.54 & 8.83 & 2.53 \\
    \hline
	\hline
	\end{tabular}
	\endgroup
\end{table*}

\section{Results}\label{sec:results}
\subsection{Evolutionary outcomes}\label{sec:triple-evolution}
After generating our initial conditions as described above, we evolve the systems forward in time until one of the following  outcomes is achieved:
\begin{itemize}
\item[(i)] The inner orbit is disrupted due to a SN;
\item[(ii)] The system becomes dynamically unstable;
\item[(iii)] The tertiary companion fills its Roche lobe (tertiary RLO);
\item[(iv)] The inner binary stars merge;  
\item[(v)] The inner binary becomes a DCO and the tertiary is lost in a SN explosion;
\item[(vi)] The system becomes a stable triple in which the inner binary is a DCO. The tertiary companion can be either another compact object or a low mass star that will neither  undergo a SN nor fill its Roche lobe in its following evolution.
\end{itemize}

\begin{figure*}
\vspace{-5pt}
\centering
\includegraphics[width=0.7\textwidth]{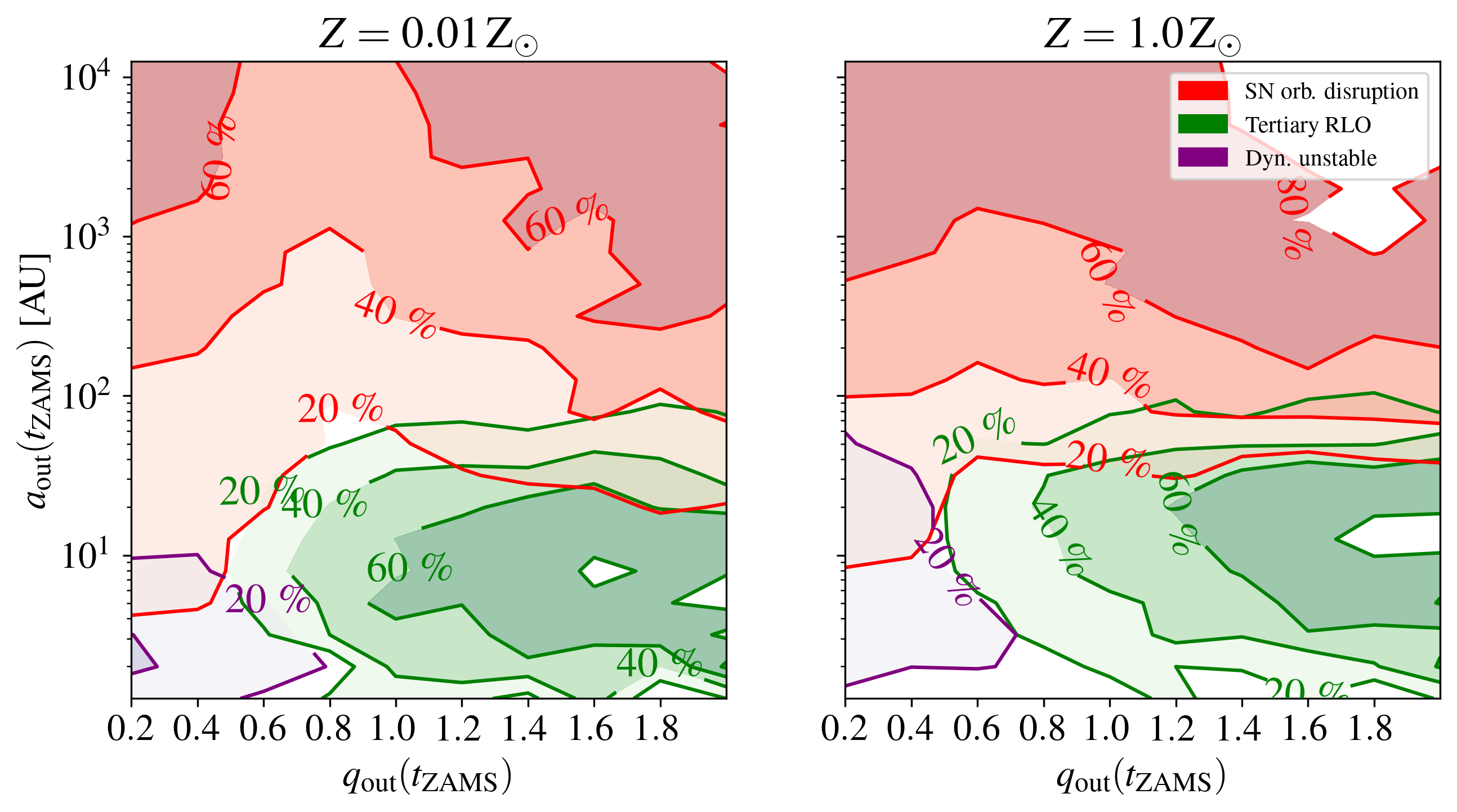}
\caption{Probability for evolutionary outcomes as a function of the initial (ZAMS) outer mass ratio $q_{\rm out}=m_3/m_{12}$ and outer semi-major axis $a_{\rm out}$ at $Z=0.01\zsun{}$ (left panel) and $Z=1.0\zsun{}$ (right panel) in the {\tt fallback kicks} model. For a given $q_{\rm out}$ and $a_{\rm out}$, the contours correspond to the fraction of triples that achieve a particular outcome.}
\vspace{-5pt}
\label{fig:plane}
\end{figure*}

In Table~\ref{tab:table} we provide the fractions of evolutionary outcomes for the different population models.
In case (iv), we consider any merger that involves a stellar component, i.e. either mergers of two stars or of a star and a compact object.
In the latter case, the compact object enters the envelope
of the companion star and sinks to its core before the envelope could be ejected. 
For case (v) and (vi), we distinguish between systems that end up harbouring a BBH, a neutron star black hole binary (NSBH), or a binary neutron star (BNS) in the inner binary.
An example system of case (vi) is shown in the left panel of Figure~\ref{fig:examples}. 

For comparison, Table~\ref{tab:table} also provides the fractions of orbital disruptions, stellar mergers, and surviving systems when no tertiary companion was included at all. This isolated inner binary population is  evolved with {\tt BSE}. We will compare in more detail the results from the binary and the triple populations in Section~\ref{sec:orbits}. 

In any of our models, we find that the majority of systems are either disrupted [case (i)] or  that the inner binary components merge [case (iv)]. Stellar mergers in triples have been extensively studied in previous work \citep[][]{2012ApJ...757...27A,2015ApJ...799..118P,2016MNRAS.460.3494S,2018A&A...610A..22T,2019ApJ...878...58S,2022A&A...661A..61T}. For example, it has been suggested that the resulting merger product could explain the observation of blue straggler stars in globular clusters \citep[][]{2009ApJ...697.1048P,2014ApJ...793..137N,2016ApJ...816...65A}. The merger process itself may give rise to a luminous red nova \citep[e.g.,][]{2016A&A...592A.134T,2017ApJ...835..282M,2017ApJ...834..107B,2019A&A...630A..75P}. It is expected that the merger star undergoes a brief phase with a bloated envelope \citep[][]{2007ApJ...668..435S,2020MNRAS.495.2796S}. If the outer orbit is sufficiently tight, it may be partially or entirely enclosed by the bloated star. This can lead to  transient phenomena as the tertiary companion plunges into the enlarged envelope \citep[][]{2016MNRAS.456.3401P,2021MNRAS.503.4276H}. Moreover, a sufficiently tight tertiary companion could co-evolve with the merger product star of the inner binary to form a bound (merging) BBH \citep[][]{2022arXiv220316544S}.

The fraction of surviving systems [i.e., cases (v) and (vi)]
depends on the kick prescription, metallicity, and the nature of the compact objects to be formed. It is the largest if {\tt no kicks} are considered and the lowest for the {\tt proportional kicks} which generally lead to the fastest kick velocities. Additionally, the number of surviving systems decreases toward solar metallicity where the stellar winds loosen the orbits and less massive remnants are formed which experience stronger natal kicks in the {\tt fallback kicks} model. Lastly, NSs experience stronger natal kicks than BHs, making the NSBH and BNS a subdominant population in the kick models compared to  BBHs. In all models, we find that the fraction of surviving DCOs that lost their tertiary companion [i.e., case (v)] is higher than those that retain it and end up as stable triples [i.e., case (vi)]. 

In Figure~\ref{fig:plane}, we plot the evolutionary outcomes of triples as a function of the initial values of $q_{\rm out}$ and $a_{\rm out}$, for the {\tt fallback kicks} model. The contours correspond to the probability that: at least one member of the triple is ejected through a SN, case (i) or (v), the system becomes dynamically unstable, case (ii), or the  tertiary undergoes RLO, case (iii), after they started from a given point in the plane.
Clearly, there is a well-defined mapping between the final evolutionary outcomes and the initial properties of the tertiary companion. 

The red contours in Figure~\ref{fig:plane} show that disruptions due to a SN occur mostly for systems with a large $a_{\rm out}$
since tertiaries on wider orbits are more easily unbound by a natal kick. 
Below $q_{\rm out}\approx0.5$, we find that more than $50\,\%$ of the systems are disrupted if $a_{\rm out}\gtrsim400\,\rm AU$. This primarily occurs due to a SN explosion in one of the inner binary components. At solar metallicity the kicks in these SNe are typically high enough to unbind both orbits. In contrast, if the inner SN occurs in a metal-poor and sufficiently hierarchical triple ($a_{\rm out}/a_{\rm in}\gtrsim10^3$), it cannot easily disrupt the compact inner binary, but only the loosely bound outer orbits by sufficiently shifting the inner binary centre of mass. Above $q_{\rm out}\gtrsim0.5$, disruptions occur primarily due to a SN explosion of the initially most massive tertiary companion which unbinds the outer orbit while leaving the inner orbit bound.

The purple contours in  Figure~\ref{fig:plane} represent systems that become dynamically unstable according to Eq.~\eqref{eq:stability}.
While reaching this regime is achieved or facilitated by the expansion of the inner orbit due to stellar winds from metal-rich binary members or, more rarely, due to a non-disruptive SN, a large number of systems at both metallicities become unstable during a Roche lobe overflow in the inner binary. Typically, the first phase of Roche lobe overflow is initiated by the primary star which expands more rapidly than its secondary companion. During the subsequent mass transfer phase, the inner binary mass ratio inverts, allowing $a_{\rm in}$ to grow by a factor $\sim\mathcal{O}(1)$ \citep[][]{2006epbm.book.....E}. Thus, triples with a close tertiary companion (preferentially $a_{\rm out}\lesssim10\,\rm AU$) become dynamically unstable, leading to a chaotic evolution in which the ejection or collision of the stars is likely. An example of this evolution is presented in the middle panel of Figure~\ref{fig:examples}.
 
The green contours in  Figure~\ref{fig:plane}
represent systems in which the tertiary companion fills its Roche lobe according to Eq.~\eqref{eq:Roche-out}. An example case is shown in the right panel of Figure~\ref{fig:examples}. In general, this occurs when the tertiary companion is close ($a_{\rm out}\lesssim10^2\,\AU$) and relatively massive ($q_{\rm out}\gtrsim0.5$). Outside that parameter region, the radius of the tertiary star is either too small to fill its Roche lobe, the inner binary becomes unstable, or undergoes a collision before the tertiary star fills its Roche lobe. The subsequent evolution of the inner binaries might be significantly affected by the mass donated by the tertiary star. For instance, if the inner binary stars become compact objects,  it is expected that accretion will increase and equalise the component masses leading to a reduced merger time and, if present, transforming a NS into a BH \citep[][]{2020MNRAS.493.1855D}. If the tertiary mass transfer is unstable, a common-envelope encompassing all three components will drain a large amount of energy and angular momentum of the orbits and allow for a diverse set of outcomes, including the merger of the inner binary and a chaotic evolution leading to the ejection of one component \citep[][]{2021MNRAS.500.1921G}. However, given the uncertainty related to mass transfer between just two stars, we opt for stopping the integration of systems when the tertiary fills its Roche lobe. For the {\tt fallback kicks} model, we find that $5.5\%$ ($1.2\%$) of the inner binaries at low (high) metallicity develop a BH component before the tertiary  fills its Roche lobe. Those binaries may give rise to an X-ray signal as they accrete matter from the tertiary. Meanwhile, $0.3\%$ ($0.4\%$) developed a NS. 

In summary, a stellar triple has to circumvent a number of defeating events in order to form a stable triple with an inner DCO. Those events demarcate distinct regions in the orbital parameter space. Most frequently, the triples are either disrupted by strong natal kicks or due to a stellar merger that takes place in the inner binary. In the following section, we will focus on the orbital properties of the surviving systems.

\subsection{Orbital properties of the surviving systems}\label{sec:orbits}
In this section, we investigate the properties of  systems in which the inner binary becomes a DCO [case (v) and (vi) above]. 

{In Table~\ref{tab:table-triples}, we give the fraction of surviving triples which are accompanied by a low-mass star and those in which the tertiary is a compact object. In any model, the number of BBHs in the inner binary which are accompanied by another BH is roughly equal or dominate those with a low-mass star by a factor of four to five. 
No surviving triple was found with a NS in the outer orbit. 
Table~\ref{tab:table-triples} gives those systems in which the tertiary is still dynamically relevant at the end of the simulation  and could possibly affect the following evolution of the inner DCO through the LK mechanism.
At low metallicity, we find that the tertiary perturbation is suppressed by the inner binary's Schwarzschild precession, i.e., $\pi t_{\rm 1PN}/j_{\rm in}t_{\rm LK}\lesssim1$, in a significant portion of the triples, e.g. $46\,\%$ in the {\tt fallback kicks} model. At solar metallicity, almost all surviving triples ($88\,\%$ in the {\tt fallback kicks} model) have a dynamically important tertiary.}
Interestingly, in the models in which we apply a finite kick to the compact objects we find no triples with an inner NS component and in which the tertiary is still dynamically relevant. We conclude that the LK mechanism is unlikely to produce any compact object binary merger in which one of the inner components is a NS.

\begin{figure*}
\vspace{-5pt}
\centering
\includegraphics[width=0.8\textwidth]{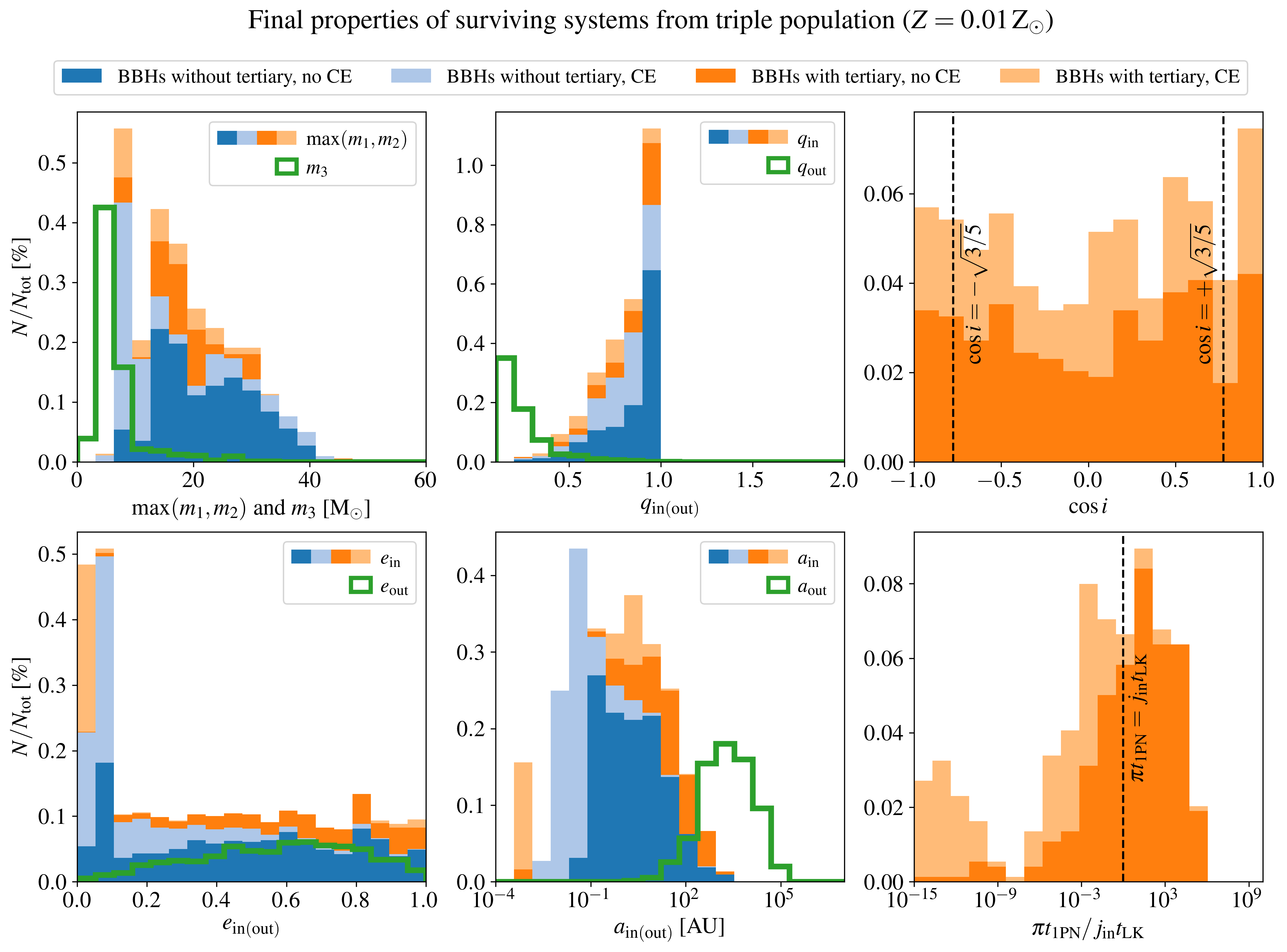}
\caption{Final orbital properties of surviving systems after a BBHs has formed in the inner binary in the {\tt fallback kicks} model. By that time, the orange systems are still accompanied by a tertiary which is either a compact object or a low mass star and whose properties are shown in green. The blue BBHs have lost their tertiary companion. For both groups we show the inner binaries that undergo and survive a CE using light colours. Blue and orange contributions are stacked.}
\vspace{-5pt}
\label{fig:survivors-1}
\end{figure*}

\begin{figure*}
\vspace{-5pt}
\centering
\includegraphics[width=0.8\textwidth]{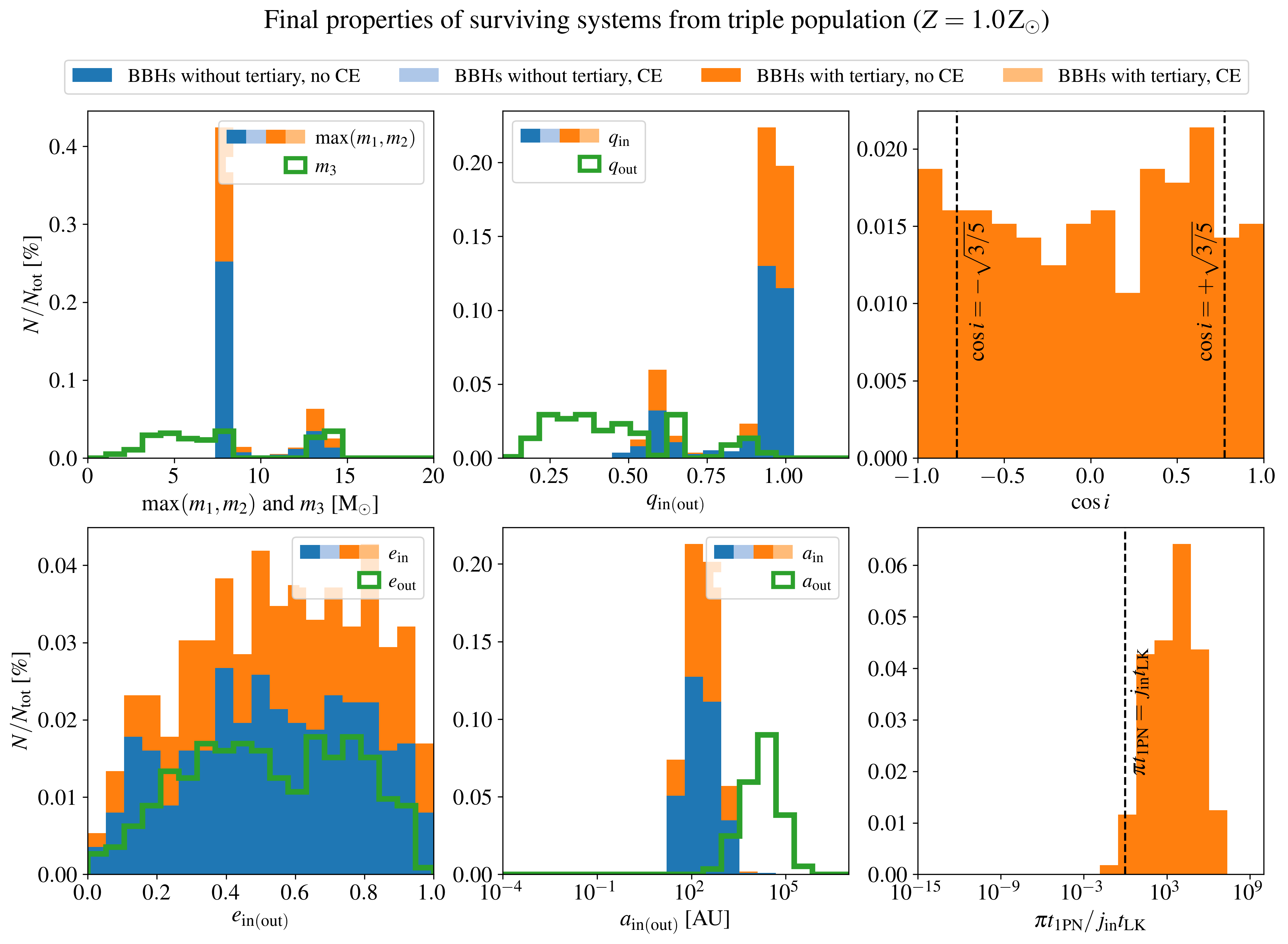}
\caption{Same as Figure~\ref{fig:survivors-1} for $Z=1.0\zsun$.}
\vspace{-5pt}
\label{fig:survivors-100}
\end{figure*}

In Figures~\ref{fig:survivors-1} and~\ref{fig:survivors-100}, we plot the orbital parameters of the surviving systems in our models for $Z=0.01\,\zsun$ and $1.0\zsun$ in the {\tt fallback kicks} model, respectively. We distinguish between DCOs which are either still accompanied by a tertiary low mass star or compact object (orange histograms), or which end up isolated (blue histograms). In either case, the large majority of inner binaries are BBHs (see Table~\ref{tab:table}).

At $Z=0.01\zsun$, the mass distribution of the primary component of the inner binary (upper left panel) peaks at $\simeq 20\msun$ and extends to $\simeq 40\msun$. 
The cut-off at $\simeq 40\msun$ is partly because we adopted an initial maximum component mass of $100\msun$. Extending the initial mass function above this mass value is unlikely to  significantly change the overall shape of the mass distributions because such massive stars are very rare. Moreover, pair-instablity SN will suppress the formation of BHs more massive than $50^{+20}_{-10}\msun$ \citep[e.g.,][]{2016A&A...594A..97B,2017MNRAS.470.4739S,10.1093/mnras/stx2933}. 

At solar metallicity, $Z=1.0\zsun$, the primary mass distribution is significantly different. Stronger wind-mass loss prior to BH formation suppresses the formation of BHs with a mass above about $15\msun$ \citep{2012ApJ...749...91F,2015MNRAS.451.4086S}. The pronounced peak at $8\msun$ primarily comes from BHs formed by $25$~--~$35\msun$  stars and initially more massive stars ($45$~--~$60\msun$) which lost additional mass in some mass transfer episode. A secondary peak at $13\msun$ relates to initially very massive stars ($\gtrsim80\msun$) which remain detached from their companion.

At both metallicities, the resulting mass ratio distribution shows a clear preference for equal masses, $q_{\rm in}\approx 1$, but otherwise differ significantly. At solar metallicity, the mass distribution of the secondary BH also shows two peaks at $8\msun$ and $13\msun$. Consequently, the mass ratio shows a secondary peak at $q\approx8\msun/13\msun\approx0.6$. In contrast, both BH component masses at low metallicity follow a much broader distribution leading to a smooth decrease of mass ratios down to $q_{\rm in}\approx0.3$.

Compared to the parent distributions (see Figure~\ref{fig:MdS}), the inner and outer semi-major axes of the surviving triples are significantly changed because  of { systems  that become dynamically unstable or merge}, and by inner binary interactions, and at high metallicity by stellar winds. At both metallicities, a large fraction of inner binaries  are prone to merge during stellar evolution and, if they are accompanied by a nearby tertiary star, to be removed due to dynamical instability or a tertiary RLO. Nonetheless, small values $a_{\rm in}\lesssim10^{-1}\,\rm AU$ are recovered in the metal-poor population because of systems in which the inner binary semi-major axis shrinks due to  a CE phase, leading to a final distribution with approximately the same median value $\bar{a}_{\rm in}\approx1$--$2\,\AU$  as the initial distribution. At solar metallicity instead, the vast majority of inner binaries that  undergo a CE phase merge. Moreover, the orbital expansion driven by the stronger stellar winds shifts the inner semi-major axis of surviving systems to  higher values, with a median $\bar{a}_{\rm in}\approx200\,\AU$. Likewise, the final value of $a_{\rm out}$ is on average larger than its initial value  due to the removal of close tertiaries which induce dynamical instability or fill their Roche-lobe and due to stellar winds of metal-rich stars. As a result, the medians of $a_{\rm out}$ increase from an initial $\sim500\,\rm AU$  to $\sim2\times10^3\,\rm AU$ and $\sim2\times10^4\,\rm AU$ at $Z=0.01\zsun$ and $Z=1.0\zsun$, respectively.
  
We find that $21\,\%$ of the surviving triples at $Z=0.01\zsun$ experience a phase of CE evolution prior to the formation of the inner DCO. At solar metallicity this is the case for none of the survivors. The zero fraction of systems that survive a CE phase at high metallicity is caused by the rapid expansion of metal-rich stars in the HG that initiate a CE, leading to stellar mergers due to the absence of a well-developed core-envelope structure. In contrast, metal-poor stars remain relatively compact in the HG but expand more significantly in the subsequent stellar evolution \citep[][]{2020A&A...638A..55K}. Consequently, a larger fraction of donor stars at lower metallicities initiate a CE during the post-HG evolution which allows for successful envelope ejection.
The efficient inspiral and circularisation during a CE phase leads to low values of $a_{\rm in}$ and $e_{\rm in}$, although a small residual eccentricity can be attained during a second SN. This type of evolution produces two characteristic  features in the distributions shown in Figures~\ref{fig:survivors-1}: the peak near $e_{\rm in}\approx0$ seen in the bottom-left panel; and the presence of DCOs at relatively small semi-major axis value, $a_{\rm in}\lesssim 1\rm AU$.
As a consequence of the decreasing $a_{\rm in}$, we find that
$\pi t_{\rm 1PN}/j_{\rm in}t_{\rm LK}<1$ for most of these triples, as shown in the bottom-right panel.
Thus, the dynamical influence of the tertiary is expected to be fully negligible for the subsequent evolution of virtually all DCOs formed from  binaries that experience a CE phase.

Regarding the DCOs that lost their tertiary companion (blue histograms in Figures~\ref{fig:survivors-1} and~\ref{fig:survivors-100}), we find a much larger fraction that underwent a CE evolution and end up at relatively low values of $a_{\rm in}$ and $e_{\rm in}$ compared to the triples that retain their companion. 
We use Eq.~\eqref{eq:t-GW} to compute the fraction of isolated DCO mergers. Based on the orbital properties at the time when the DCO is formed, we find $2.2\,\%$ ($0.16\,\%$) BBHs, $0.04\,\%$ ($0.03\,\%$) NSBHs, and $0.001\,\%$ ($0.01\,\%$) BNSs with $\tau_{\rm coal}<10^{10}\,\rm yr$ at low (high) metallicity in the {\tt fallback kicks} model. In the {\tt no kicks} model we have $3.0\,\%$ ($0.24\,\%$) BBHs, $0.14\,\%$ ($0.04\,\%$) NSBHs, and $0.14\,\%$ ($0.4\,\%$) BNSs and in the {\tt proportional kicks} $1.9\,\%$ ($0.14\,\%$) BBHs, $0.04\,\%$ ($0.006\,\%$) NSBHs, and $0.003\,\%$ ($0.017\,\%$) BNSs.

\begin{figure*}
\vspace{-5pt}
\centering
\includegraphics[width=0.8\textwidth]{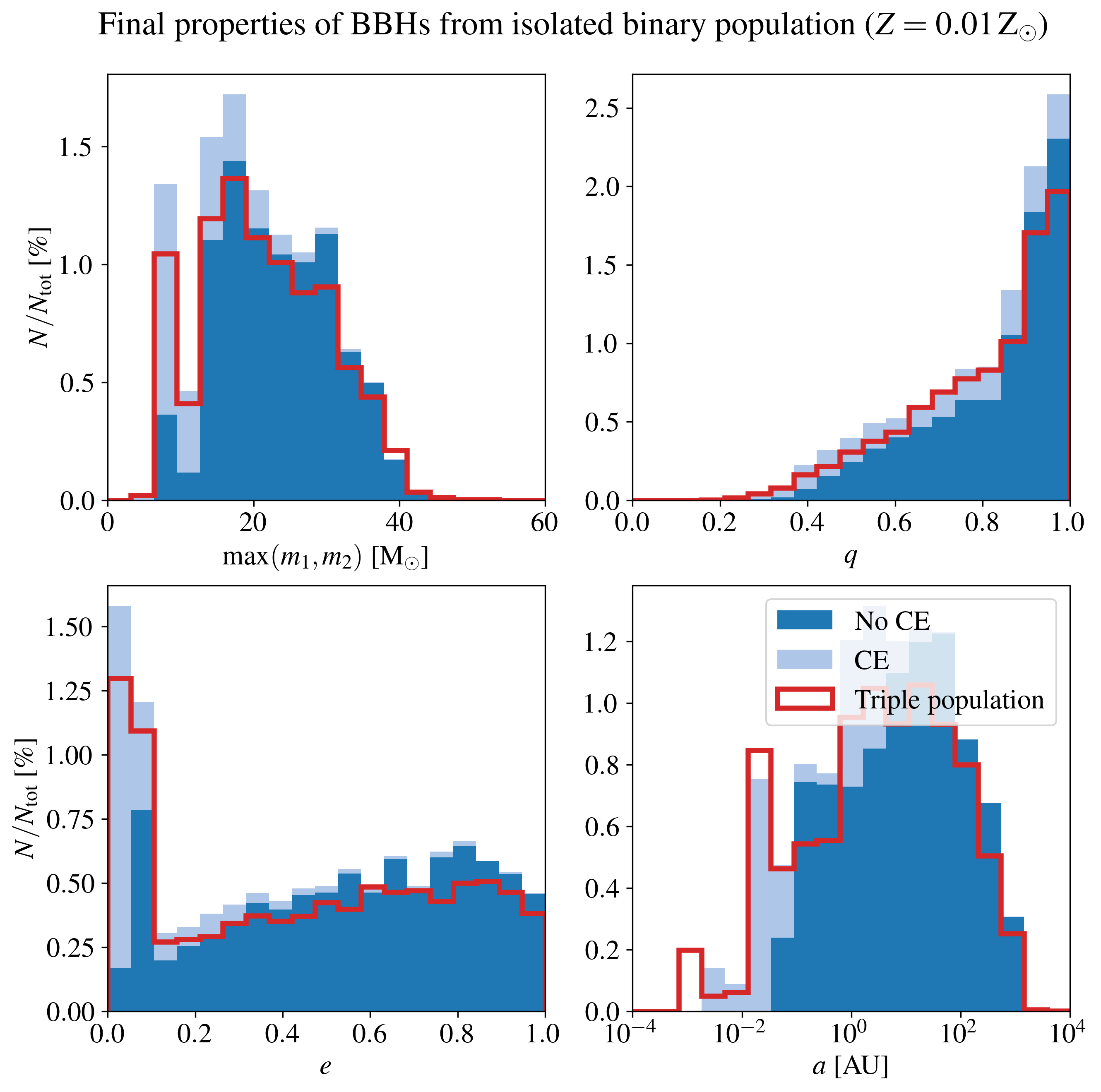}
\caption{Final orbital properties of BBHs which are formed from an isolated binary population. The population is initially identical to the inner binaries of our triples. We distinguish between binaries that undergo and survive a common-envelope evolution (CE) and those which do not (no CE). For comparision, we also show the distribution of the inner BBHs that form in the triple population in the {\tt fallback kicks} model (red), cf. Figure~\ref{fig:survivors-1}.}
\vspace{-5pt}
\label{fig:isolated-dcos-1}
\end{figure*}

\begin{figure*}
\vspace{-5pt}
\centering
\includegraphics[width=0.7\textwidth]{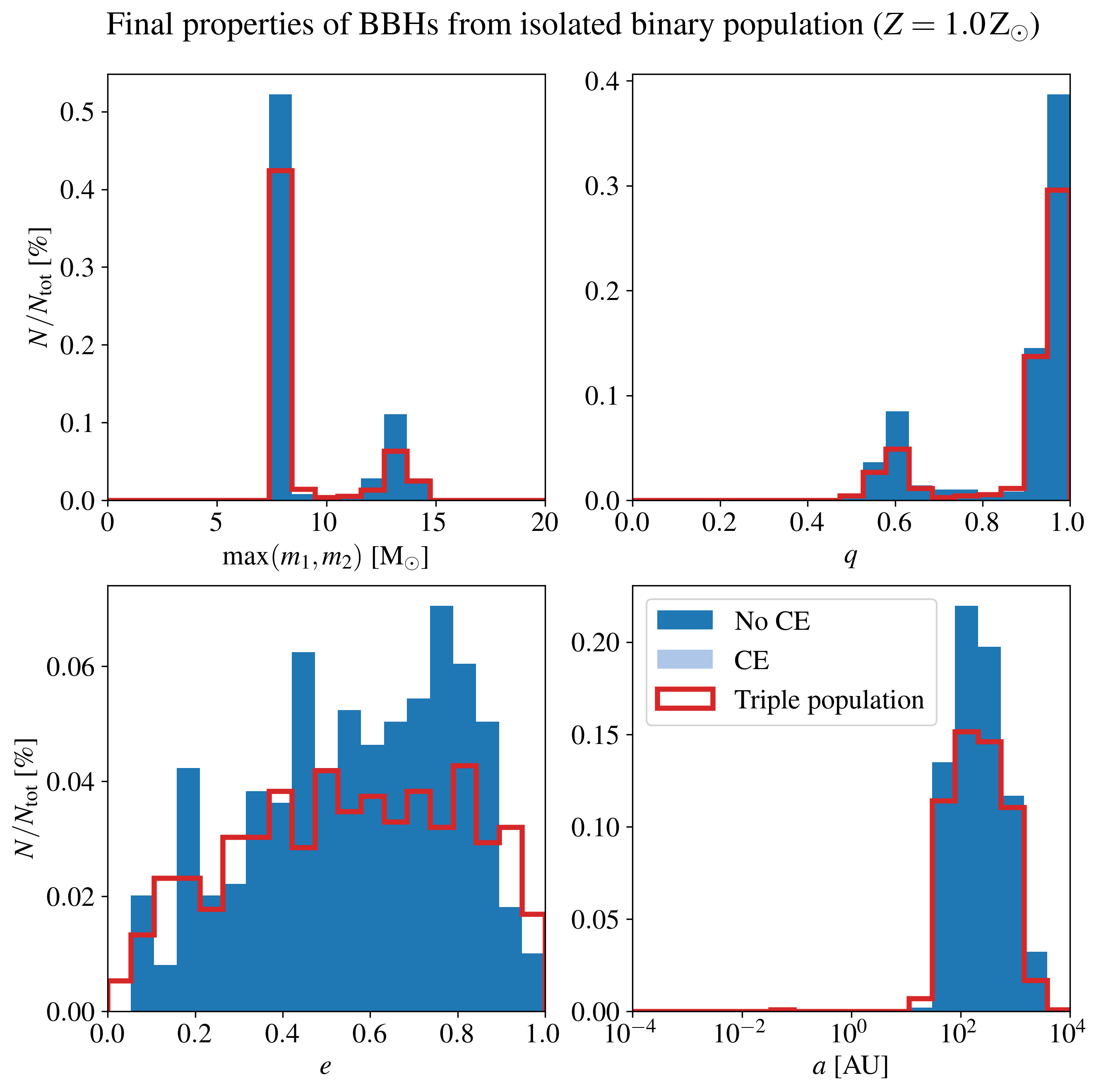}
\caption{Same as Figure~\ref{fig:isolated-dcos-1} for $Z=1.0\zsun$.}
\vspace{-5pt}
\label{fig:isolated-dcos-100}
\end{figure*}

It is useful to compare the distribution of all DCOs formed in the triple population to those that formed from an equivalent isolated binary population, i.e. binaries that evolve without an outer companion from the beginning. To this end, we evolve the same inner binaries of our triple population with {\tt BSE} and give the fractions of different evolutionary outcomes in Table~\ref{tab:table}. Figures~\ref{fig:isolated-dcos-1} and~\ref{fig:isolated-dcos-100} show the orbital properties of the DCOs in the two populations for the  {\tt fallback kicks} model.
Overall, the number of surviving DCOs from the triple population is smaller due to systems that become dynamically unstable or whose integration is terminated due to a tertiary RLO. Yet, the overall shape of the parameter distributions is similar. 
{Likewise, in the other kick models we find no significant differences between the shape of the parameter distributions between the binary and triple population models.}
This suggests that the presence of a tertiary companion does not significantly affect the final orbital distribution of the DCOs formed in our models. 




\begin{table*}
	\centering
	\caption{Detailed fraction of surviving triples for our different models at sub-solar and solar metallicity. The number of systems harbouring a DCO in the inner binary as reported in Table~\ref{tab:table} are further refined by distinguishing between triples in which the tertiary companion is a low-mass star ("+Star") and a BH ("+BH"). There is no surviving triple with a NS tertiary. The numbers in parentheses indicate the fractions of systems which are LK-possible in the sense that $\pi t_{\rm 1PN}>j_{\rm in}t_{\rm LK}$ at the end of the simulation.}
	\label{tab:table-triples}
	\begingroup
	\renewcommand*{\arraystretch}{1.2}
	\begin{tabular}{cccccccc}
	\hline
	\hline
	\multirow{2}{*}{Z $[\rm Z_\odot]$} & \multirow{2}{*}{Model} & \multicolumn{6}{c}{Fraction of evolutionary outcomes $N/N_{\rm tot}$ [$\%$]}\\
	& & $\Gamma_{\rm BBH+Star}$ & $\Gamma_{\rm BBH+BH}$ & $\Gamma_{\rm NSBH+Star}$ & $\Gamma_{\rm NSBH+BH}$ & $\Gamma_{\rm BNS+Star}$ & $\Gamma_{\rm BNS+BH}$\\
    \hline
	\hline
    \multirow{4}{*}{0.01} & {\tt Fallback kicks} & 0.72 (0.29) & 2.84 (1.65) & 0.03 (0.00) & 0.02 (0.00) & 0.00 (0.00) & 0.00 (0.00) \\
    & {\tt Proportional kicks} & 0.15 (0.01) & 0.14 (0.03) & 0.02 (0.00) & 0.00 (0.00) & 0.00 (0.00) & 0.00 (0.00) \\
    & {\tt No kicks} & 1.00 (0.45) & 4.27 (2.57) & 0.39 (0.27) & 0.74 (0.59) & 0.05 (0.00) & 0.05 (0.02) \\
    & {\tt Incl. dyn. tides} & 0.56 (0.27) & 2.97 (1.74) & 0.00 (0.00) & 0.03 (0.01) & 0.00 (0.00) & 0.00 (0.00) \\
    \hline
    \multirow{4}{*}{1.0} & {\tt Fallback kicks} & 0.14 (0.13) & 0.12 (0.10) & 0.00 (0.00) & 0.00 (0.00) & 0.00 (0.00) & 0.00 (0.00) \\
    & {\tt Proportional kicks} & 0.00 (0.00) & 0.00 (0.00) & 0.00 (0.00) & 0.00 (0.00) & 0.00 (0.00) & 0.00 (0.00) \\
    & {\tt No kicks} & 0.34 (0.33) & 1.40 (1.27) & 0.72 (0.68) & 0.66 (0.62) & 0.08 (0.03) & 0.06 (0.02) \\
    & {\tt Incl. dyn. tides} & 0.13 (0.11) & 0.16 (0.13) & 0.00 (0.00) & 0.00 (0.00) & 0.01 (0.00) & 0.00 (0.00) \\
    \hline
    \hline
	\end{tabular}
	\endgroup
\end{table*}

\subsection{Tertiary impact on inner binary interactions}\label{sec:stellar-interactions}
We previously identified certain regions of parameter space where the tertiary companion  induces dynamical instability, is ejected by a SN, or overflows its Roche lobe. In this section, we investigate how the companion affects the evolution of the inner binary stars. It is well-known that massive stars in binaries are prone to closely interact and undergo one or more episodes of mass transfer \citep[][]{1967AcA....17..355P,1992ApJ...391..246P,2012Sci...337..444S,2013ApJ...764..166D,2016A&A...588A..10R,2021PhRvD.103f3007S,2021MNRAS.507.5013M}. Here, we determine whether the interaction with a tertiary companion changes the stellar evolution of the inner binary stars and the overall fraction of systems that experience a mass transfer phase.

\begin{figure*}
\vspace{-5pt}
\centering
\includegraphics[width=0.8\textwidth]{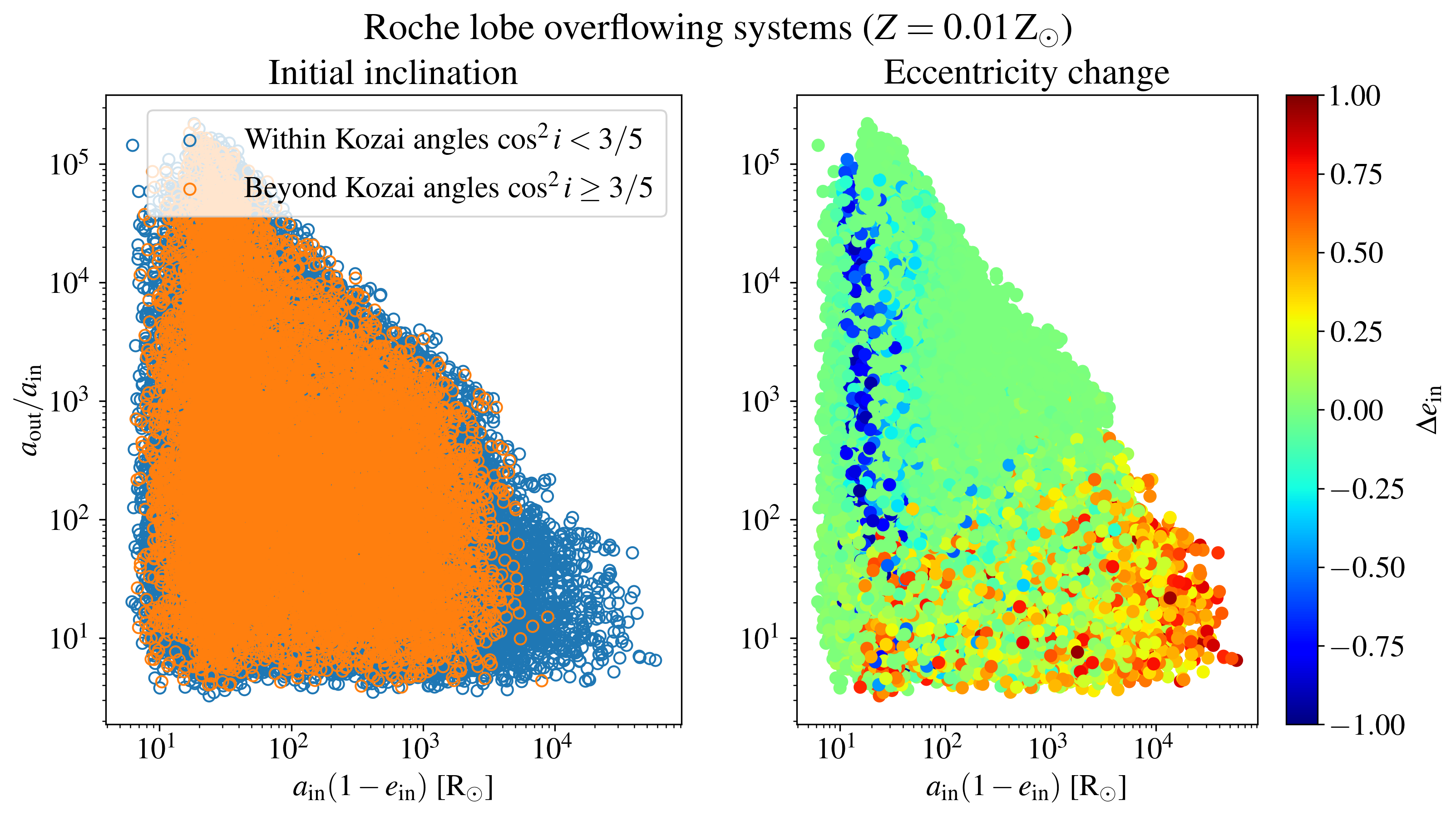}
\includegraphics[width=0.8\textwidth]{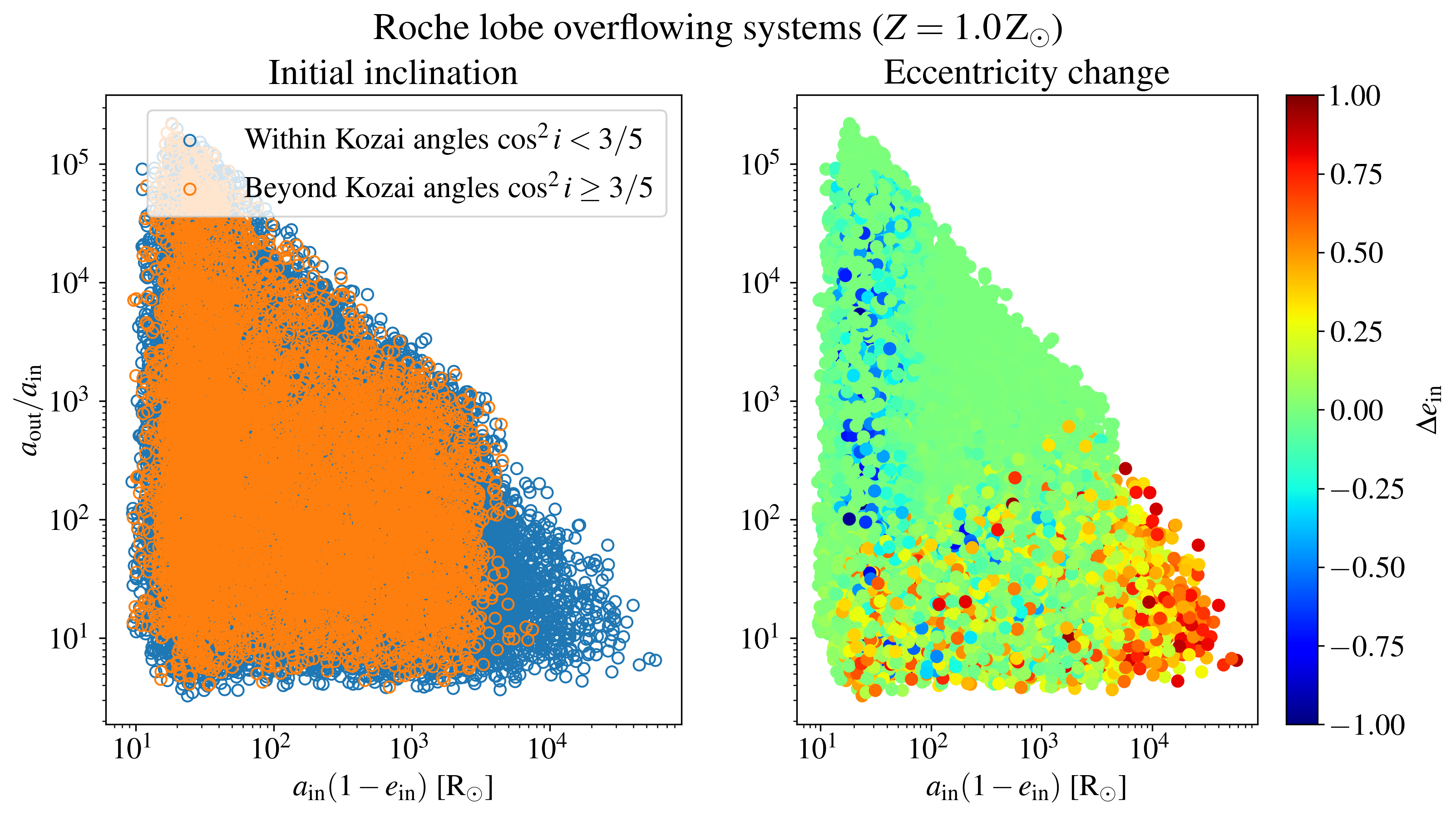}
\caption{All triples whose inner binaries undergo a phase of mass transfer at $Z=0.01\zsun$ (upper panels) and $Z=1.0\zsun$ (lower panels). Plotted are the initial values for their semi-major axis ratio $a_{\rm out}/a_{\rm in}$ and inner periapsis $a_{\rm in}(1-e_{\rm in})$. The colour scheme on the left panels indicates the initial relative inclination between the inner and outer orbital plane. On the right panels, we indicate the eccentricity change $\Delta e_{\rm in}$ between the initial time and the onset of mass transfer.}
\vspace{-5pt}
\label{fig:FRLO}
\end{figure*}

In Figure~\ref{fig:FRLO}, we plot the initial distribution of the semi-major axis ratio $a_{\rm out}/a_{\rm in}$ and periapsis $a_{\rm in}(1-e_{\rm in})$ of triples in which the inner binaries undergo a phase of mass transfer (either stable or unstable). This is the case for $83\,\%$ ($87\,\%$) of all systems at $Z=0.01\,\rm Z_\odot$ ($Z=1.0\,\rm Z_\odot$). In the left panels, we highlight whether the initial relative inclination is in the LK angle regime ($\cos^2 i<3/5$) and in the right panels, we show the differences between the initial eccentricity and its value at the onset of mass transfer.\footnote{If an inner binary undergoes multiple mass transfer phases we are considering the first one.} We note that Roche lobe overflow outside the Kozai angle regime ($\cos^2 i\geq3/5$) only occurs if the initial periapsis is below $a_{\rm in}(1-e_{\rm in})\lesssim 10^3\,\rm R_\odot$ 
However, if the relative inclination is within the LK angle, Roche lobe overflow is possible in initially wider orbits and up to $\lesssim10^5\,\rm R_\odot$. In these systems, the tertiary companion excites the inner eccentricity via LK oscillations and effectively reduces the periapsis so that the stars have to be less expanded in order to fill their Roche lobe. Roche lobe overflow in those inner binaries is therefore induced by the perturbation from the tertiary companion. 

In the right panels of Figure~\ref{fig:FRLO} we show the change in eccentricity between the initial time and the onset of mass transfer.
For $a_{\rm in}(1-e_{\rm in})\gtrsim 10^3\,\rm R_\odot$ the binary eccentricity is  higher than its initial value ($\Delta e_{\rm in}>0$), demonstrating the impact of the LK mechanism. 
A considerable fraction of $16.4\,\%$ ($16.4\,\%$) of Roche lobe overflowing systems at $a_{\rm in}(1-e_{\rm in})<10^3\,\rm R_\odot$, also have a  higher eccentricity ($\Delta e_{\rm in}>0.1$). Furthermore, these binaries are found to fill their Roche lobe at an earlier evolutionary stage than in an equivalent run without tertiary companion.
This shows that the impact of the LK mechanism extends to essentially all values of $a_{\rm in}$, but only for semi-major axis ratios below $a_{\rm out}/a_{\rm in}\lesssim10^2$.
Finally, if $a_{\rm in}(1-e_{\rm in})\lesssim\mathcal{O}(10^1)\,\rm R_\odot$, 
the binary orbits can be significantly affected by tides. These binaries circularise due to tidal friction ($\Delta e_{\rm in}<0$).
Similar results are found by \citet[][]{2020A&A...640A..16T} who considered less massive triples with initial primary masses $1$~--~$7.5\,\rm M_\odot$.

Lastly, we investigate whether the tertiary companion changes the fraction of binaries that experience a specific type of close interaction. More specifically, we distinguish between four types of stellar interactions:
\begin{itemize}
\item[(i)] The inner binary stars merge;
\item[(ii)] The two stars do not merge, but undergo and survive at least one phase of CE;
\item[(iii)] The binary neither merges nor experiences a CE phase, but 
 undergoes at least one phase of stable mass-transfer;
\item[(iv)]
If none of cases (i)~--~(iii) applies, the inner binary will evolve without undergoing any strong interaction and the stars will effectively  behave as if they were single stars. Thus, we refer to this latter type  of evolution as ``effectively single''. 
\end{itemize}

\begin{figure*}
\vspace{-5pt}
\centering
\includegraphics[width=0.7\textwidth]{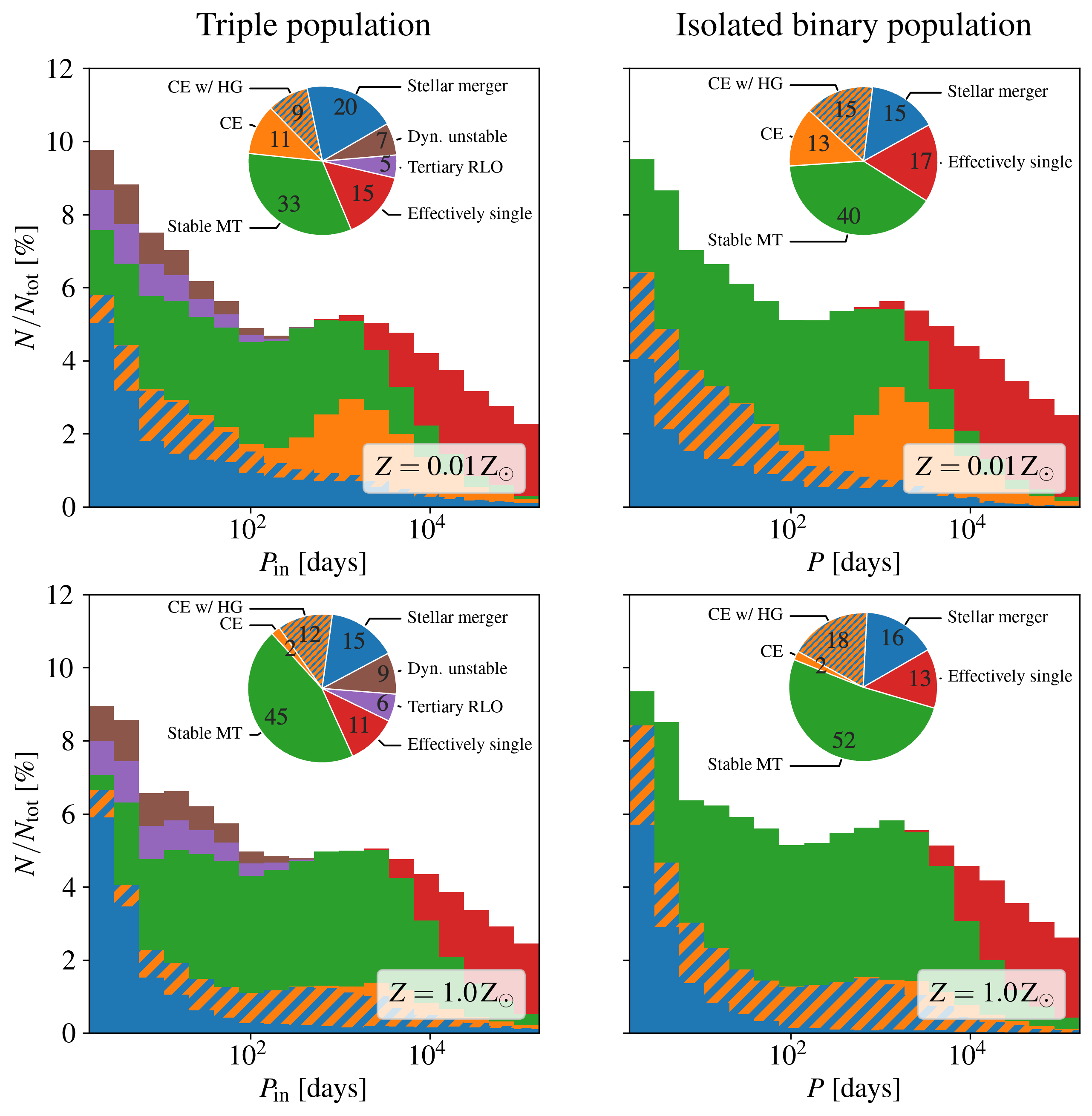}
\caption{Fractions of systems in a triple population (left panels) and isolated binary population (right panels) that undergo a certain kind of close stellar interaction as a function of their initial (inner) orbital period in the {\tt fallback kicks} model.}
\vspace{-5pt}
\label{fig:triple-interactions}
\end{figure*}

As in Section~\ref{sec:results}, merger refers to any coalescence of the inner binary that involves at least one stellar component. In the left panels of Figure~\ref{fig:triple-interactions}, we show the fraction of binary interactions for $Z=0.01$ (upper panel) and $1.0\,\zsun{}$ (lower panel) as a function of their initial inner orbital period. Evidently, close stellar interactions between the massive inner binary members are prevalent at both metallicities since only $15\%$ and $12\%$ of them evolve as effectively single stars for $Z=0.01$ and $1.0\,\zsun{}$, respectively. The type of interaction depends on the binary orbital period. At $P_{\rm in}\lesssim10\,\rm days$, the vast majority of inner binary stars merge. For those we highlight the binaries that merge in a common-envelope which is initiated by a donor in the HG. Toward longer orbital periods the fraction of binary stars which undergo a stable mass transfer episode increases until the population becomes dominated by stars that do not interact at all (above $P_{\rm in}\gtrsim10^4\,\rm days$). The major difference between the two metallicities lies in the fraction of systems that survive a CE phase (around $P_{\rm in}\approx10^3\,\rm days$), which is $10\%$ at $Z=0.01\,\zsun$ and only $2\%$ at $Z=1.0\,\zsun$. 

Systems whose evolution is terminated due to a tertiary Roche lobe overflow or due to dynamical instability are shown separately and found at short periods $P_{\rm in}\lesssim10^2\,\rm days$ (cf. Section~\ref{sec:triple-evolution}). Together these system contribute $12\,\%$ ($15\,\%$) at $Z=0.01\zsun$ ($Z=0.01\zsun$). Although their evolution is uncertain, we should expect that the triple interaction will leave a significant imprint on the evolution of the stars in these systems.
In dynamical unstable systems, one member (typically the lightest star) is likely to be ejected from the triple leaving a bound pair of stars behind \citep[]{2001MNRAS.321..398M}. 
For tertiary Roche lobe overflow  it can be expected that the inner binary will undergo some sort of interaction during the subsequent evolution, which is further perturbed by the mass accreted from the tertiary \citep[][]{2020MNRAS.491..495D,2020MNRAS.493.1855D,2021MNRAS.500.1921G,2021arXiv211000024H}.

In the right panels of Figure~\ref{fig:triple-interactions}, we show the same analysis for the binary population model in which the initially identical inner binaries are evolved without tertiary companion (see Table~\ref{tab:table}). The phenomenon of tertiary-induced interactions as discussed in the previous section amounts to a decrease by less than $3\,\%$ of effectively single inner binaries in the triple population. Hence, the presence of a tertiary companion only marginally changes the number of systems that evolve as effectively single binaries. On the other hand, as discussed above, the inner binary evolution is more significantly affected at short orbital periods, where we see systems that undergo a tertiary RLO or become dynamical unstable.

\begin{figure*}
\vspace{-5pt}
\centering
\includegraphics[width=0.7\textwidth]{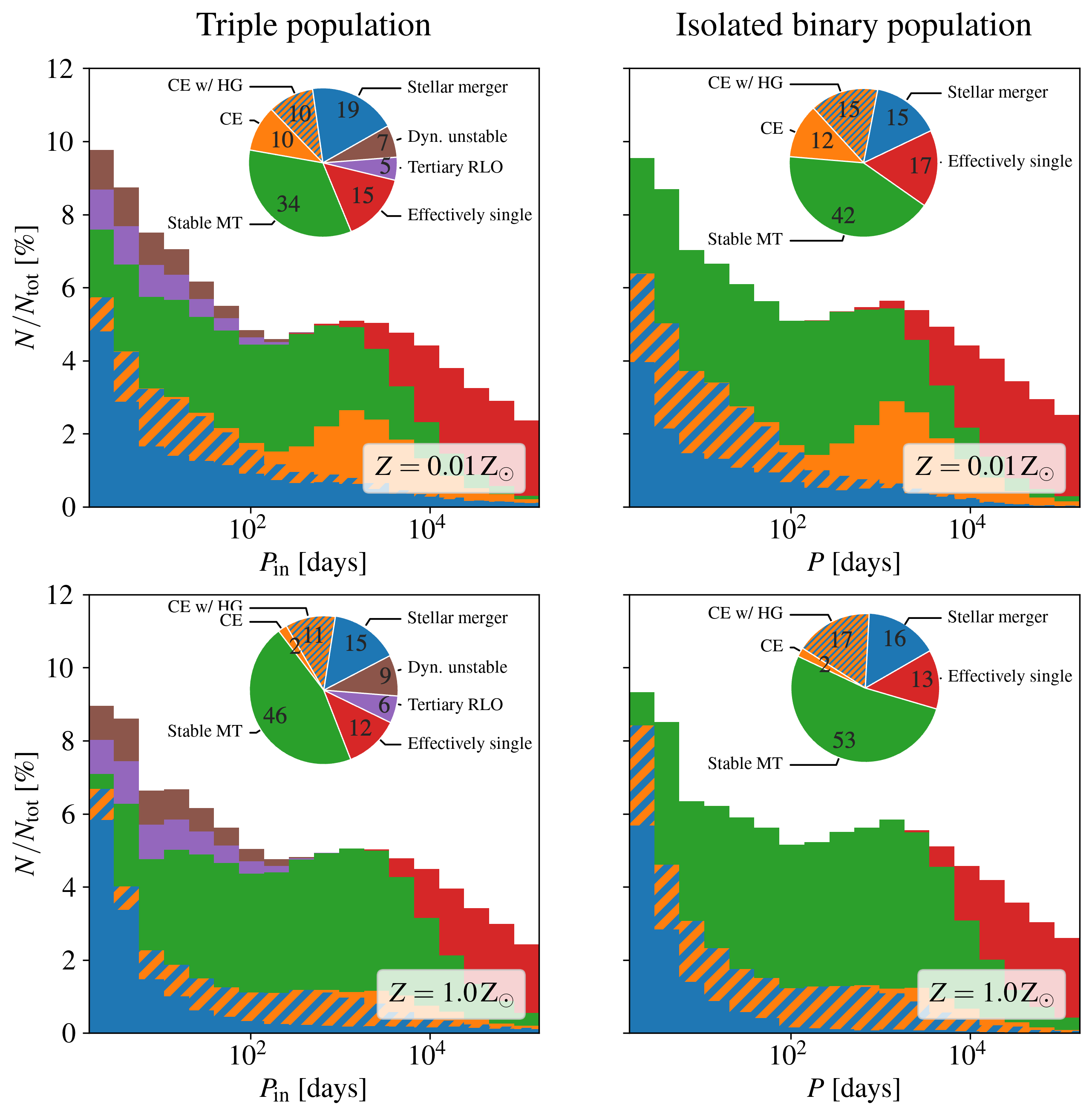}
\caption{Same as Figure~\ref{fig:triple-interactions} in the {\tt proportional kicks} model.}
\vspace{-5pt}
\label{fig:triple-interactions-proportional}
\end{figure*}

\begin{figure*}
\vspace{-5pt}
\centering
\includegraphics[width=0.7\textwidth]{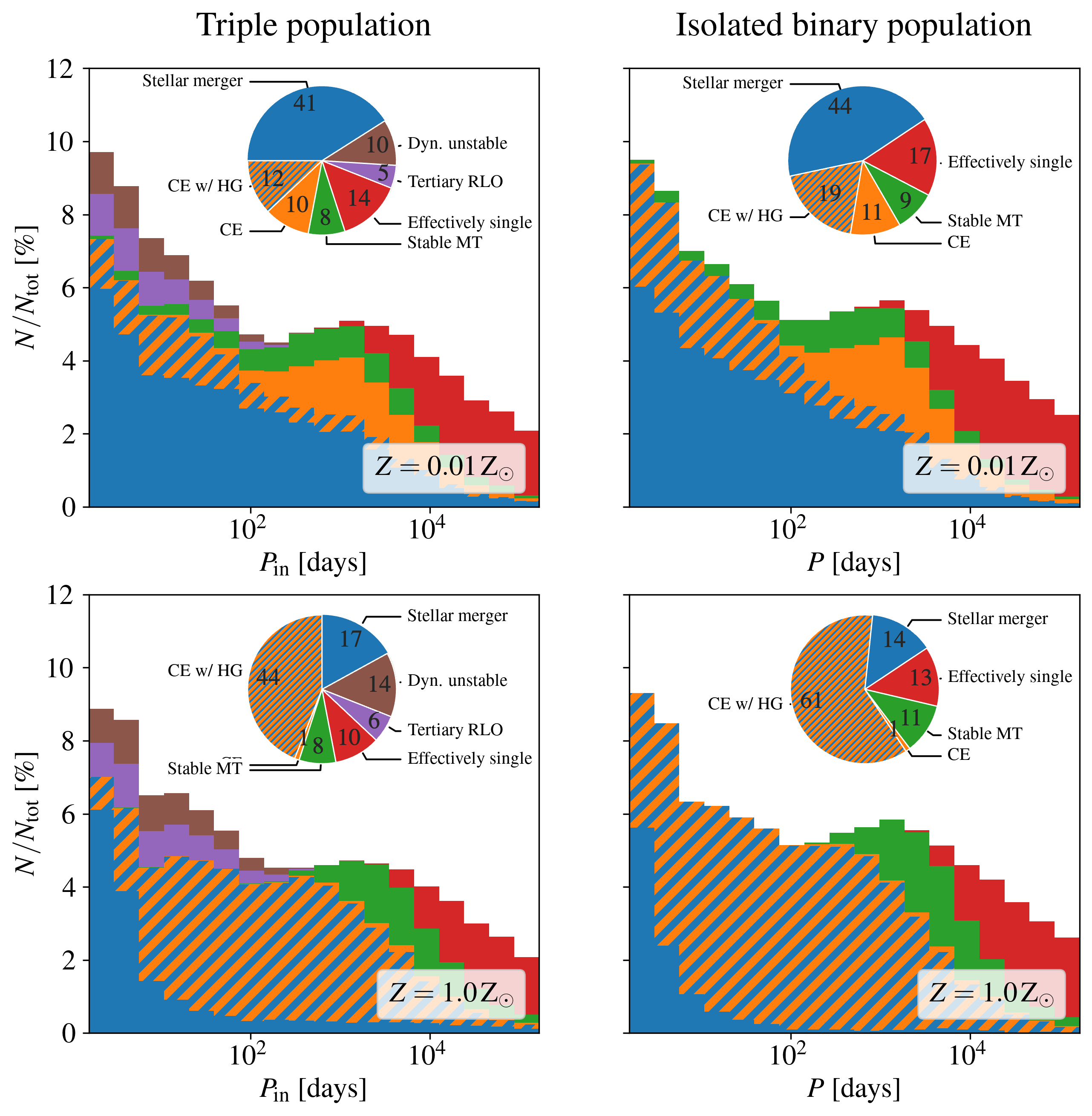}
\caption{Same as Figure~\ref{fig:triple-interactions} in the {\tt no kicks} model.}
\vspace{-5pt}
\label{fig:triple-interactions-no}
\end{figure*}

In Figures~\ref{fig:triple-interactions-proportional} and~\ref{fig:triple-interactions-no}, we show the same comparison in the {\tt proportional kicks} and {\tt no kicks} model, respectively. While the  former is nearly identical to the {\tt fallback kicks} model, the latter shows a much higher fraction of systems which merge or undergo a common-envelope evolution with a donor in the HG. These happen when the binary companion is already a compact object. In the non-zero kick models, these systems  tend to be disrupted already at the formation of the compact object due to a natal kick.

\section{Conclusions}\label{sec:conclusions}
In this work, we used our code {\tt TSE} to study the long-term evolution of a massive stellar triple population starting from the ZAMS  until they form compact objects. {\tt TSE} simultaneously takes into account the secular three-body interaction and the evolution of the stars.  In the following, we summarise and discuss the main results of our work.

\begin{enumerate}

\item There is a well-defined mapping between the initial properties of the triples and the probability to achieve a certain evolutionary outcome (Figure~\ref{fig:plane}). Wide systems are vulnerable to disruption by a SN kick. In our models with non-zero kicks we find that in more than $50\,\%$ of triples with an outer semi-major axis $a_{\rm out}\gtrsim 400\,\rm AU$ the inner binary is disrupted during a SN.
At smaller values of  $a_{\rm out}$, most systems either 
experience a merger in the inner binary before a DCO is formed, become dynamically unstable (for $q_{\rm out}\lesssim 0.8$), or  have tertiary companion that fills its Roche lobe (for $q_{\rm out}\gtrsim 0.8$).
Stellar mergers can give rise to observable signatures such as red novae \citep[][]{2019A&A...630A..75P} and hydrogen-rich \citep[][]{2019ApJ...876L..29V} or strongly magnetised remnants \citep[][]{2019Natur.574..211S}. Dynamically unstable systems enter a regime in which our secular approach breaks down and a chaotic evolution takes place leading to the ejection of one component or the merger of two \citep[][]{2001MNRAS.321..398M,2015ApJ...808..120P,2022A&A...661A..61T}. The subsequent evolution of systems with a Roche lobe filling tertiary companion is not well understood. Compared to RLO in isolated binaries, previous work on tertiary RLO is inherently more complex due to modulations caused by the periodic motion of the inner binary and its non-trivial response to mass accretion \citep[][]{2020MNRAS.491..495D,2020MNRAS.493.1855D,2021MNRAS.500.1921G,2021arXiv211000024H}.

\item Our method provides a self-consistent way to generate compact object triples (and DCOs with a low mass star companion) which can be used to study the triple channel for gravitational wave sources \citep[][]{2017ApJ...836...39S,2018MNRAS.480L..58A,2017ApJ...841...77A,2018ApJ...863....7R}.  Table~\ref{tab:table} gives the fraction of  triple evolutionary outcomes for our different models, showing that at most a few percent of systems evolve to become stable triples with an inner DCO binary --  the exact fraction depends on metallicity and the adopted natal kick prescription.
The orbital properties of all  systems that form a DCO binary are shown in Figures~\ref{fig:survivors-1} and~\ref{fig:survivors-100}.
At low metallicity, more than half of the surviving triples are LK-possible in the sense that LK oscillations are not quenched by the Schwarzschild precession of the inner binary orbit. At solar metallicity, this is the case for almost all triples.  

\item In any of our models, the vast majority of surviving triples harbour an inner BBH (see Table~\ref{tab:table-triples}). Triples with a NS component in the inner binary are very rare. In  models with non-zero natal kicks  their number is typically two orders of magnitude smaller than triples with BBHs. Unless {\tt no kicks} are assumed, no surviving triple harbouring a BNS  has been found in our population. We conclude that it is unlikely that BNSs can be driven to a merger via the LK mechanism in triples. 

\item Population synthesis studies of massive stellar binaries do not follow the  
interaction of the binary with outer companions. However, treating the inner binary 
as isolated
poses a risk to miss out the perturbative effect a tertiary companion on the evolution of the inner binary.
{Our study shows that inner binaries with initial periapses $10^3\lesssim a_{\rm in}(1-e_{\rm in})/\rm R_\odot\lesssim10^5$ are driven to a RLO due to the presence of the tertiary companions (Figure~\ref{fig:FRLO}). The latter can effectively reduce the minimum periapses so that the inner binary stars have to be less expanded in order to fill their Roche lobe. This gives rise to mass transfer episodes on very eccentric orbits. Below $\sim10^3\,\rm R_\odot$ the inner binary stars undergo RLOs even if they were in isolation. Nonetheless, the tertiary companions can cause them to occur on more eccentric orbits provided that $a_{\rm out}/a_{\rm in}\lesssim10^2$ initially.}

\item By comparing the triple population to isolated binary population models, we show that the interaction with the tertiary companion does not significantly change the resulting orbital distributions of the surviving (inner) DCOs. Moreover, in the triple population the fraction of systems in which the inner binaries evolve without undergoing a mass transfer episode is only decreased by not more than $3\,\%$ compared to the binary population models (Figures~\ref{fig:triple-interactions}~--~\ref{fig:triple-interactions-no}). However, compact triples with initial inner periods $P_{\rm in}\lesssim10^2\,\rm days$ are prone to become dynamically unstable or to have a Roche lobe filling tertiary companion. We find this to be the case in $\sim7$~--~$14\,\%$ and $\sim5\,\%$ of the systems, respectively (Table~\ref{tab:table}). For these systems the evolution of the inner binary is expected to be significantly affected by the triple interaction.

\end{enumerate}

\section*{Acknowledgements}
We acknowledge the support of the Supercomputing Wales project, which is part-funded by the European Regional Development Fund (ERDF) via Welsh Government. For the numerical simulations we made use of {\fontfamily{qcr}\selectfont GNU Parallel} \citep{tange_ole_2018_1146014}. FA acknowledges support from a Rutherford fellowship (ST/P00492X/1) from the Science and Technology Facilities Council. MM acknowledges financial support from NASA grant ATP-170070\,(80NSSC18K0726).

\section*{Data Availability}
The code that implements the methods of this article is publicly available on our {\fontfamily{qcr}\selectfont GitHub} repository, \url{https://github.com/stegmaja/TSE}.



\bibliographystyle{mnras}
\bibliography{triple_population}



\appendix

\section{Hut's polynomials}\label{sec:Hut}
\begin{align}
    f_1(e_{\rm in})&=1+\frac{31}{2}e_{\rm in}^2+\frac{255}{8}e_{\rm in}^4+\frac{185}{16}e_{\rm in}^6+\frac{25}{64}e_{\rm in}^8,\\
    f_2(e_{\rm in})&=1+\frac{15}{2}e_{\rm in}^2+\frac{45}{8}e_{\rm in}^4+\frac{5}{16}e_{\rm in}^6,\\
    f_3(e_{\rm in})&=1+\frac{15}{4}e_{\rm in}^2+\frac{15}{8}e_{\rm in}^4+\frac{5}{64}e_{\rm in}^6,\\
    f_4(e_{\rm in})&=1+\frac{3}{2}e_{\rm in}^2+\frac{1}{8}e_{\rm in}^4,\\
    f_5(e_{\rm in})&=1+3e_{\rm in}^2+\frac{3}{8}e_{\rm in}^4.
\end{align}

\section{Exemplary triples}\label{appendix:examples}
The initial parameters of the three triples shown in Figure~\ref{fig:examples} are as following.

The masses of the system in the left panel are $m_1=98.4\msun{}$, $m_2=32.6\msun{}$, and $m_3=44.1\msun{}$. The eccentricities and semi-major axes are $e_{\rm in}=0.89$, $e_{\rm out}=0.48$, $a_{\rm in}=30.7\AU$, and $a_{\rm out}=915.2\AU$, respectively. The relative inclination reads $\cos i=0.26$ while the arguments of periapses are $\sin\omega_{\rm in}=0.11$ and $\sin\omega_{\rm out}=0.87$.

For the systems in the middle panel we have $m_1=60.8\msun$, $m_2=44.0\msun$, $m_3=20.2\msun$, $e_{\rm in}=0.99$, $e_{\rm out}=0.68$, $a_{\rm in}=74.9\AU$, $a_{\rm out}=1125.3\AU$, $\cos i=-0.83$, $\sin\omega_{\rm in}=-0.06$, and $\sin\omega_{\rm out}=0.72$.

Finally, the parameters of the system in the right panel are $m_1=49.6\msun$, $m_2=33.0\msun$, $m_3=40.0\msun$, $e_{\rm in}=0.83$, $e_{\rm out}=0.38$, $a_{\rm in}=0.8\AU$, $a_{\rm out}=17.2\AU$, $\cos i=0.77$, $\sin\omega_{\rm in}=0.27$, and $\sin\omega_{\rm out}=0.82$.

\end{document}